\documentclass[10pt,twocolumn,twoside]{IEEEtran}
\IEEEoverridecommandlockouts

\usepackage{cite}
\ifCLASSINFOpdf
  \usepackage[pdftex]{graphicx}
\else
  \usepackage[dvips]{graphicx}
\fi
\usepackage[cmex10]{amsmath}
\usepackage{amssymb}
\usepackage{algorithm,algorithmic}
\floatname{algorithm}{Protocol}
\usepackage{color}

\usepackage{bm}
\usepackage{bbm}

\newcommand{\mathskip}{\hspace{0.2cm}}
\newcommand{\bigmathskip}{\hspace{0.5cm}}

\renewcommand{\vec}{\bm}
\newcommand{\mathset}{\mathcal}
\newcommand{\mat}{\bm}

\newtheorem{remark}{Remark}

\floatname{algorithm}{Algorithm}

\usepackage{pgfplots}
\pgfplotsset{compat=1.8}
\usepackage{tikz}
\usetikzlibrary{calc,intersections,through,backgrounds,shapes.geometric,decorations.pathmorphing,arrows,plotmarks,external,patterns}
\tikzsetexternalprefix{./externaltikz/}
\usepackage[absolute,showboxes]{textpos}

\usepackage[xindy,toc,nonumberlist]{glossaries}
\usepackage{glossary-mcols}
\setglossarystyle{mcolalttree}
\glssetwidest[0]{SC-FDMA}
\newacronym{ARQ}{ARQ}{automatic repeat request}
\newacronym{AM}{AM}{Acknowledged Mode}
\newacronym[description=Access Point]{AP}{AP}{access point}
\newacronym{BLER}{BLER}{Block Error Rate}
\newacronym{BP}{BP}{Backpressure}
\newacronym[description=Base Station]{BS}{BS}{base station}
\newacronym{BSR}{BSR}{Buffer Status Report}
\newacronym{CAC}{CAC}{Call Admission Control}
\newacronym{CC}{CC}{congestion control}
\newacronym{CDMA}{CDMA}{Code Division Multiple Access}
\newacronym{CE}{CE}{Control Element}
\newacronym[description=Cellular Mode]{CMode}{CM}{cellular mode}
\newacronym{CP}{CP}{Control-Plane}
\newacronym{CQI}{CQI}{Channel Quality Indicator}
\newacronym{D2D}{D2D}{Device-to-Device}
\newacronym{DFT}{DFT}{Discrete Fourier Transformation}
\newacronym[description=Downlink]{DL}{DL}{downlink}
\newacronym[description=Direct Mode]{DMode}{DM}{direct mode}
\newacronym{DP}{DP}{Data-Plane}
\newacronym{FCFS}{FCFS}{first-come first-serve}
\newacronym{FDD}{FDD}{Frequency Division Duplex}
\newacronym{FM-PCA}{FM-PCA}{Foschini-Miljanic PCA}
\newacronym{GBD}{GBD}{Generalized Benders Decomposition}
\newacronym{GSM}{GSM}{Global System for Mobile Communications}
\newacronym{HARQ}{HARQ}{Hybrid Automatic Repeat Request}
\newacronym[description=Internet of Things]{IoT}{IoT}{internet of things}
\newacronym{IP}{IP}{Internet Protocol}
\newacronym{KKT}{KKT}{Karush-Kuhn-Tucker}
\newacronym[description=Local Area Network]{LAN}{LAN}{local area network}
\newacronym{LCFS}{LCFS}{last-come first-serve}
\newacronym{LQG}{LQG}{linear-quadratic-Gaussian}
\newacronym{LTE}{LTE}{Long Term Evolution}
\newacronym{LTE-A}{LTE-A}{LTE Advanced}
\newacronym{LTE-U}{LTE-U}{LTE Unlicensed}
\newacronym{LTI}{LTI}{linear time-invariant}
\newacronym{M2M}{M2M}{machine-to-machine}
\newacronym{MAC}{MAC}{Medium Access Control}
\newacronym{MCS}{MCS}{Modulation and Coding Scheme}
\newacronym[description=Mixed-Integer Nonlinear Problem]{MILP}{MILP}{mixed-integer linear problem}
\newacronym[description=Mixed-Integer Nonlinear Problem]{MINLP}{MINLP}{mixed-integer nonlinear problem}
\newacronym{MNO}{MNO}{Mobile Network Operator}
\newacronym[description=Mobile Station]{MS}{MS}{mobile station}
\newacronym{MTC}{MTC}{Machine-Type Communication}
\newacronym{NCS}{NCS}{networked control system}
\newacronym[description=Nonlinear Problem]{NLP}{NLP}{nonlinear problem}     
\newacronym[description={Non-Deterministic, Polynomial Time}]{NP}{NP}{non-deterministic, polynomial time}
\newacronym{NR}{NR}{New Radio}
\newacronym[description=Network Utility Maximization]{NUM}{NUM}{network utility maximization}
\newacronym{OFDM}{OFDM}{Orthogonal Frequency Division Multiplex}
\newacronym{OFDMA}{OFDMA}{Orthogonal Frequency Division Multiple Access}
\newacronym[description=Power Control]{PC}{PC}{power control}
\newacronym[description=Power Control Algorithm]{PCA}{PCA}{power control algorithm}
\newacronym{PDCP}{PDCP}{Packet Data Convergence Protocol}
\newacronym{PDU}{PDU}{Packet Data Unit}
\newacronym{PF}{PF}{Proportional Fair}
\newacronym{PHY}{PHY}{Physical}
\newacronym[plural={p.p.p.'s},longplural={Poisson Point Processes}]{ppp}{p.p.p.}{Poisson Point Process}
\newacronym{PRB}{PRB}{Physical Resource Block}
\newacronym{QAM}{QAM}{Quadrature Amplitude Modulation}
\newacronym{QoS}{QoS}{Quality of Service}
\newacronym{QPSK}{QPSK}{Quadrature Phase Shift Keying}
\newacronym{RAP}{RAP}{Random Access Procedure}
\newacronym{RLC}{RLC}{Radio Link Control}
\newacronym{RM}{RM}{Reuse Mode}
\newacronym{SC-FDM}{SC-FDM}{Single-Channel Frequency Division Multiplex}
\newacronym{SC-FDMA}{SC-FDMA}{Single-Channel Frequency Division Multiple Access}
\newacronym{SINR}{SINR}{Signal to Interference and Noise Ratio}
\newacronym{SIR}{SIR}{Signal to Interference Ratio}
\newacronym{SDU}{SDU}{service data unit}
\newacronym[description=Sidelink]{SL}{SL}{sidelink}
\newacronym{SMS}{SMS}{Short Message Service}
\newacronym{TBS}{TBS}{Transport Block Sizie}
\newacronym{TCP}{TCP}{Transmission Control Protocol}
\newacronym{TDD}{TDD}{Time Division Duplex}
\newacronym{TDMA}{TDMA}{Time Division Multiple Access}
\newacronym{TM}{TM}{Transparent Mode}
\newacronym{TTI}{TTI}{Transmission Time Interval}
\newacronym{UDP}{UDP}{User Datagram Protocol}
\newacronym[description=User Equipment]{UE}{UE}{user equipment}
\newacronym[description=Uplink]{UL}{UL}{uplink}
\newacronym{UM}{UM}{Unacknowledged Mode}
\newacronym{UMTS}{UMTS}{Universal Mobile Telecommunications System}
\newacronym{WCDMA}{WCDMA}{Wideband Code Division Multiple Access}
\newacronym{WLAN}{WLAN}{Wireless LAN}
\newacronym{WSR}{WSR}{weighted sum-rate}
\glsaddall
\makenoidxglossaries

\TPMargin{5pt}

\begin{document}

%\IEEEpubid{978-1-5386-3873-6/17/ \textdollar 31.00 \textcopyright 2017 European Union}

\bstctlcite{IEEEexample:BSTcontrol}
%%
%% paper title
%% can use linebreaks \\ within to get better formatting as desired
%% Do not put math or special symbols in the title.
%\title{Bare Demo of IEEEtran.cls for Conferences}

\author{
	\IEEEauthorblockN{
		Markus Kl\"ugel\IEEEauthorrefmark{1}, 
		Mohammad H. Mamduhi\IEEEauthorrefmark{3}, 
		Onur Ayan\IEEEauthorrefmark{1}, 
		Mikhail Vilgelm\IEEEauthorrefmark{1},
		Karl H. Johansson\IEEEauthorrefmark{3}\\
		Sandra Hirche\IEEEauthorrefmark{2}, 
		Wolfgang Kellerer\IEEEauthorrefmark{1}
	}
	\IEEEauthorblockA{
		\IEEEauthorrefmark{1}Chair of Communication Networks, Technical University of Munich\\
		\IEEEauthorrefmark{2}Chair of Information-Oriented Control, Technical University of Munich\\
		\IEEEauthorrefmark{3}Division of Decision and Control Systems, KTH Royal Institute of Technology, Stockholm\\
		{$\lbrace$markus.kluegel, onur.ayan, mikhail.vilgelm, hirche, wolfgang.kellerer$\rbrace$@tum.de, $\lbrace$mamduhi, kallej$\rbrace$@kth.se}
	}

\thanks{This work has been funded by the German Research Foundation (DFG) under the grant number 315177489 as part of the SPP 1914 (CPN).}
}


%Telephone: (800) 555--1212, Fax: (888) 555--1212}


%\copyrightstatement
%\title{Joint Cross-layer Optimization of Multi-loop Control and Multi-hop Network in Networked Control Systems}
%\title{Joint Cross-layer Optimization in Networked Control Systems}
\title{Joint Cross-layer Optimization in Real-Time Networked Control Systems}
%\title{Cross-layer Optimal Co-design of Control and Network in Multi-hop Networked Control Systems}
%\title{Joint Network Optimization and Controller Design for Networked Control Systems}
%\title{A Decomposition Approach to Networked Control Systems}
\maketitle

\begin{abstract}
 \Gls{NCS} refer to a set of control loops that are closed over a communication network. In this article, the joint operation of control and networking for \gls{NCS} is investigated wherein the network serves the sensor-to-controller communication links for multiple stochastic \gls{LTI} sub-systems. The sensors sample packets based on the observed plant state, which they send over a shared multi-hop network. The network has limited communication resources, which need to be assigned to competing links to support proper control loop operation. In this set-up, we formulate an optimization problem to minimize the weighted-sum \gls{LQG} cost of all loops, taking into account the admissible sampling, control, congestion control and scheduling policies. Under some mild assumptions on the sampling frequencies of the control loops and the communication network, we find the joint optimal solution to be given by a certainty equivalence control with threshold-based sampling policy, as well as a back-pressure type scheduler with a simple pass-through congestion control. The interface between network and control loops is identified to be the buffer state of the sensor node, which can be interpreted as network price for sampling a packet from control perspective. We validate our theoretical claims by simulating NCSs comprising of multiple LTI stochastic control loops communicating over a two-hop cellular network.
\end{abstract}


% \fi
\IEEEpeerreviewmaketitle
\glsresetall
\section{Introduction}

\vspace{-1.5mm}
\subsection*{Motivation}
\Gls{M2M} and \gls{IoT} are envisioned as driving, revenue-generating applications for the near future of communication networks. They include a wide range of applications in vertical domains, e.g., smart grids, vehicular communications and industrial automation. While current networks were primarily designed to support high-rate, human-driven applications such as video streaming, web-browsing, or file transfer, current research focuses on wider range of heterogeneous requirements from both human- and machine-driven applications. Many \gls{M2M} applications involve communicating sensors, actuators or in general control loops that are closed over a network. Studies show the resulting control performance within these applications is tightly coupled with performance of the communication system. However, the exact relationships is non-trivial and not yet fully understood.

To efficiently support \gls{M2M} applications, the interplay between control performance and the underlying communication system capabilities has to be precisely studied. %In an attempt to understand it, \gls{NCS} have been introduced as a research topic, and have gained much attention from academia and industry. 
In this line of work until very recently, two rather independent perspectives have been dominant among the control and communication societies: while from control perspective, the communication network capabilities are typically abstracted as maximum rate, delay and packet loss properties, parallel approaches from the communication community abstract control applications by their requirements on rate, delay and packet loss. This leads eventually to a separate design of control and communication yet ignoring their non-trivial coupling. It is shown in a variety of recent works that \textit{joint design of communication and control systems} in \gls{NCS} provides flexible networking algorithms and improves control performance.

\vspace{-2mm}

\subsection{Contributions}
In this work, we investigate optimal joint design of network and control strategies by minimizing the weighted sum \gls{LQG} cost of multiple stochastic \gls{LTI} control loops that share a communication network. We tackle the task in a cross-layer fashion \cite{4118456}, optimizing over all possible sampling and control policies, as well as over \gls{CC} and scheduling strategies. We use a generalized system model that allows application to a variety of networks, in particular to wireline ethernet, cellular, ad-hoc networks and satellite communication. The results can be applied to single-hop or multi-hop networks. To the best of our knowledge, the combination of multi-loop control with multi-hop networks is novel; further we are among the first to consider \gls{NCS} in an interdisciplinary fashion of this depth and generalizability. We show ways for interaction between network and control beyond explicit rate and delay constraints, which we see as critical point for ensuring the right level of compatibility and decoupling among both disciplines. To limit the delay effect, we focus on real-time communication in which data transport latency is negligible from control perspective due to communication sampling periods being much finer than those of control loops. The remaining problem still poses challenges~as network resources need to be traded off among the control loops, whereas each loop needs to adapt to the offered network resources.Our major contributions are:
\begin{itemize}
	\item Performance optimization of \gls{NCS} consisting of stochastic LTI systems over a shared network in cross-layer fashion over all sampling, control, CC and scheduling policies.
	\item Identification of buffer status as a natural interface between network and control loops.
		\item  Applying non-trivial decomposition methods to show that certainty equivalence control, threshold-based sampling policy, and back-pressure scheduling achieve optimality.
\end{itemize}
%The contributions are obtained using a general network model to support a large variety of communication systems, including wireline ethernet, cellular, ad-hoc and satellite networks.
\vspace{-2mm}

\subsection{Related Work}
\label{sec:relatedwork}

The problem of joint communication and control design in \gls{NCS} has been an intensive research subject in the control community. There exist two general approaches to consider communication in \glspl{NCS}: Treat given protocol and medium as constraints~\cite{bommannavar2008optimal} or consider transmissions as an additional cost~\cite{molin2009lqg}. The former approach is more common as it is aligned with the layering principle of system design \cite{4118456}, however the latter is more powerful in terms of joint optimization.

From the control side, the optimal design is studied in~\cite{molin2013optimality}, in particular it is shown under which conditions the certainty equivalence controller is optimal. Stability and performance of feedback control under delay and packet dropouts are discussed in~\cite{branicky2002scheduling, heemels09networked}. In~\cite{bommannavar2008optimal} an optimal control strategy with constraint on resources and packet loss is developed, whereas~\cite{Molin2014} studies optimal control when a joint transmission constraint enforces a trade-off among networked control loops. Authors in \cite{8405590} also proposed an LQG optimal delay-dependent sampling when usage of communication resources is costly.

From the communication side, centralized approaches for resource allocation in \glspl{NCS} have been developed in~\cite{xiao2003joint, 6161215, mamduhi2015robust}. The authors in~\cite{xiao2003joint, 6161215} consider rate scheduling, whereas~\cite{mamduhi2015robust} deal with user scheduling problems. A well-known approach to user scheduling, Maximum Error First (MEF) Try-once-Discard, has been presented in~\cite{walsh01scheduling, walsh02stability}. The authors propose to greedily schedule the control sub-system with the highest control error first. An extension of the MEF approach using finite horizon model-based prediction is introduced in~\cite{schoeffauer2018predictive}.

Several works exist for design of decentralized \gls{MAC} protocols, i.e., CSMA/CA- or ALOHA-based protocols~\cite{mamduhi2015decentralized, vilgelm2016adaptive, gatsis2016control,BALAGHII201858}, with applications to \gls{WLAN} systems~\cite{Boggia2008, Misra2015, eisen2018control}. In addition, data link layer for NCS has been studied for optimal power allocation~\cite{quevedo2010energy}, and modulation and coding schemes choice~\cite{liu2004wireless}.

Given the complexity of the joint design of NCS, state-of-the-art typically restricts the scenario to single-hop networks and a specific \gls{MAC} layer \cite{molin2009lqg,6286997}. Other communication layers or multiple hops are not considered in the optimization, however, performance evaluation case studies with multi-hop networks are available in the literature, e.g.,~\cite{li2016wireless}. 

In this article we consider a multi-loop \gls{NCS} supported~by~a multi-hop network and study the joint optimization over the sampling and control laws associated with control, as well as the CC and scheduling laws of the communication network. Our framework is applicable to both decentralized and centralized \gls{MAC} protocols, including ethernet, cellular networks, ad-hoc networks and satellite communication. While building on existing works, e.g., \cite{Molin2014}, our results are significantly more general. Further, we quantify a natural interface between control and networking, that are given a priori in related works.

\vspace{-2mm}

\subsection{Outline}
In the remainder of this article, NCS model and optimization preliminaries are presented in Section~\ref{sec:systemmodel}. Problem statement and solution are explained in Section~\ref{sec:problemstatement}. Simulation results are shown in Section~\ref{sec:simulation} and the article is concluded in Section~\ref{sec:conclusions}.

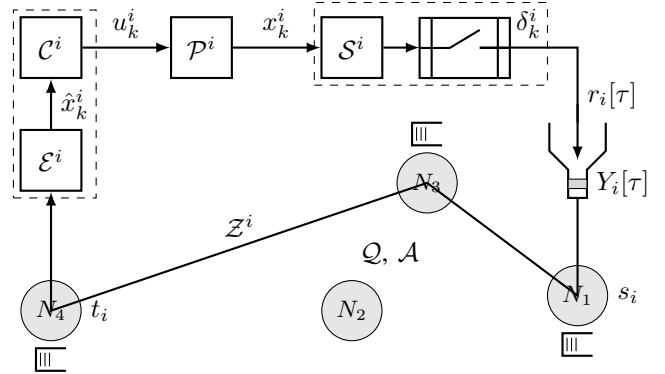
\begin{figure}
	\centering
%	\resizebox{\columnwidth}{!}{}
	\definecolor{TUMBlau}{RGB}{0,101,189} % Pantone 300
\definecolor{TUMBlauDunkel}{RGB}{0,82,147} % Pantone 301
\definecolor{TUMBlauHell}{RGB}{152,198,234} % Pantone 283
\definecolor{TUMBlauMittel}{RGB}{100,160,200} % Pantone 542

% Hervorhebung:
\definecolor{TUMElfenbein}{RGB}{218,215,203} % Pantone 7527 -Elfenbein
\definecolor{TUMGruen}{RGB}{162,173,0} % Pantone 383 - Grün
\definecolor{TUMOrange}{RGB}{227,114,34} % Pantone 158 - Orange
\definecolor{TUMGrau}{gray}{0.6} % Grau 60%

\tikzset{
		my_line/.style={
			line width= 0.8pt, % 0.4 is the default
			solid,
			arrows=-
			},
		control_line/.style={
			line width=0.6pt,
			arrows=->,
			dashed
			},
		control_box/.style={
			draw,
			my_line,
			rectangle,
			minimum height=23pt,
			minimum width=23pt,
			anchor=center,
			fill=white
		},
		double_control_box/.style={
		draw,
		my_line,
		rectangle,
		minimum height=46pt,
		minimum width=23pt,
		anchor=center,
		fill=white
		},
		radiation/.style={{my_line, decorate,decoration={expanding waves,angle=30,segment length=6pt}}},		
	}
	\tikzset{Networknode/.style = {circle,minimum width = 0.1cm,draw = black, fill=black!10!white,font=\small}}

	\newcommand{\drawQueue}[1]{
		\coordinate (Queue_Input) at (0,0.75);
		\draw[thick] (0,0) -- (2,0) -- (2,1.5) -- node[anchor = west]{#1} (0,1.5);
		\draw[] (1.1,1.2) -- (1.1,0.3);
		\draw[] (1.4,1.2) -- (1.4,0.3);	
		\draw[] (1.7,1.2) -- (1.7,0.3);
		\coordinate (Queue_Output) at (2,0.75);
	}

	\newcommand{\drawSampler}{
		\tikzset{sampler/.style = {thick}};
		\draw[sampler] (0,0) rectangle (1.5,1);
		\draw[sampler] (0.2,0) -- (0.2,1) (1.3,0) -- (1.3,1);
		\draw[sampler] (0,0.5) -- (0.5,0.5) -- (1,0.8) (1,0.5) -- (1.5,0.5);
	}
	
	\newcommand{\drawCongestionControl}[2]{
		\coordinate (CC_Input) at (1,2.5);
		\draw[thick,-latex] (1,2.5) -- node[near start,anchor=south west]{#1} (1,1);
		\draw[thick] (0.25,2) -- (0.25,1.25) -- (0.75,0.75) -- (0.75,0) -- (1.25,0) -- (1.25,0.75) -- (1.75,1.25) -- (1.75,2);
		\draw[thin,fill=gray!30!white] (0.75,0.25) rectangle (1.25,0.5);
		\node at (1.25,0.375) [anchor=west]{#2};
		\coordinate (CC_Output) at (1,0);
	}
	
\begin{tikzpicture}
%\centering

%    \draw[help lines](0,0) grid[step=1]  (9,6);
	
	\draw[black, dashed] (-1.5,3) rectangle (-0.4,5.5);
	
	\node[control_box] (C) at (-1,5) {$\mathcal{C}^i$};
	\node[control_box] (E) at (-1,3.5) {$\mathset{E}^i$};	
%	\node[control_box_thin] (E) at (-1,5) {$\mathcal{\epsilon}^i$};
	\node[control_box] (P) at (1,5) {$\mathcal{P}^i$};
	\node[control_box] (S) at (3,5) {$\mathcal{S}^i$};
	\draw[black, dashed] (2.5,4.5) rectangle (5.6,5.6);
	
	\begin{scope} [xshift = 3.9cm,yshift = 4.595cm,scale=0.8]
		\drawSampler
	\end{scope}

	\begin{scope}[xshift = 5.5cm, yshift = 3cm,scale=0.5]
		\drawCongestionControl{$r_i[\tau]$}{$Y_i[\tau]$};
	\end{scope}
%	
%	\begin{scope}[xshift = 5.625cm, yshift = 4cm,scale=0.5,rotate=-90]
%		\drawQueue{$Y_1^i[\tau]$};
%	\end{scope}
	
	\node(N1) at (6,1.7)[Networknode]{$N_1$};
	\node(N2) at (3,1.5)[Networknode]{$N_2$};
	\node(N3) at (4,3.2)[Networknode]{$N_3$};
	\node(N4) at (-1,1.5)[Networknode]{$N_4$};
	
	\draw[my_line, -latex] (C.east) -- node[anchor=south]{$u_k^i$} (P.west);	
	\draw[my_line, -latex] (P.east) -- node[anchor=south]{$x_k^i$}(S.west);
	\draw[my_line, -latex] (S.east) -- (3.9,5);
	\draw[my_line, -latex] (E.north) -- node[anchor=west]{$\hat{x}_k^i$}(C.south);	
	
	\draw[my_line, -] (4.7,5) -- node[anchor=south] {$\delta_k^i$} (6,5) -- (6,4);

	\draw[my_line, -latex] (6,3) -- (N1.center) -- (N3.center) -- node[anchor=south]{$\mathset{Z}^i$} (N4.center) -- (E.south);
	\node at (3.5,2.25) {$\mathcal{Q}$, $\mathcal{A}$};
	\node at (N1.east) [anchor=west] {$s_i$};
	\node at (N4.east) [anchor=west] {$t_i$};
	
	\begin{scope}[xshift=4.2cm,yshift=4cm,scale=0.2,rotate=180]
		\drawQueue{};
	\end{scope}
	\begin{scope}[xshift=-0.8cm,yshift=1cm,scale=0.2,rotate=180]
		\drawQueue{};
	\end{scope}
	\begin{scope}[xshift=6.2cm,yshift=1.2cm,scale=0.2,rotate=180]
		\drawQueue{};
	\end{scope}

	\end{tikzpicture}
	\vspace{-3mm}\caption{Schematic of the problem setup with one control loop and a four-node multi-hop network. The decision variables include sampling, congestion control, transmission control and feedback control.}
	\label{fig:setup}
	\vspace{-3.5mm}
\end{figure}

\section{System Model \& Preliminaries}
\label{sec:systemmodel}
Consider a multiple-loop NCS as schematically shown in Fig. \ref{fig:setup}. It consists of $L$ control loops, with the set of all loops denoted by $\mathset{L}\!=\!\{1,\ldots,L\}$, where sensor-to-controller links are closed over a multi-hop communication network.

\vspace{-2mm} 

\subsection{Control Model}
Loop $i\!\in \!\mathset{L}$ consists of a physical process $\mathset{P}^i$, sensor $\mathset{S}^i$ and a control unit including an estimator $\mathset{E}^i$ and a feedback controller $\mathset{C}^i$. All processes follow LTI dynamics disturbed by exogenous stochastic inputs. The process $\mathset{P}^i$ is described by%in discrete time by the following inherent system dynamics
\begin{equation}\label{eq:loop_dynamics}
	\vec{x}_{k+1}^i = \mat{A}^i\vec{x}_k^i + \mat{B}^i\vec{u}_k^i + \vec{w}_k^i,
\end{equation}
where $\vec{x}_k^i\!\in\!\mathbb{R}^{n_i}$ is the system state of loop $i$ at time-step $k$, $\vec{u}_k^i\!\in\!\mathbb{R}^{m_i}$ is the control input, and $\mat{A}^i\!\in\!\mathbb{R}^{n_i\times n_i}$, $\mat{B}^i\!\in\!\mathbb{R}^{n_i\times m_i}$ represent the system and control matrices, respectively. The exogenous disturbance $\vec{w}_k^i\!\in \!\mathbb{R}^{n_i}$ takes random values at each time $k$ according to a zero-mean Gaussian distribution with covariance $\mat{Z}^i$ and is assumed to be an i.i.d. process for all $i\!\in\!\mathset{L}$ and $k\in\{0,1,\ldots\}$. We, moreover, assume that the initial values $\vec{x}_0^i$'s, $i\in\mathset{L}$, are i.i.d. from distributions with symmetric density functions around their respective means $\mathbbm{E}[\vec{x}_0^i]$. The disturbances $\vec{w}_k^i$'s are presumed to be independent from the initial states $\vec{x}_0^i$'s, for all $i$ and all $k$. We assume that all control loops $i\!\in\!\mathset{L}$ evolve in discrete time with sampling periods $T^i$, i.e., time-step $k$ refers to the time instant $k\cdot T^i$ for the $i^{\text{th}}$ loop. 

Quality of control for the $i^{\text{th}}$ control loop is measured in infinite time-horizon, by the following LQG cost function $J^i$:
\begin{equation}\label{eq:individual_cost}
J^i = \limsup_{K\to\infty}\frac{1}{K}\mathbbm{E}\left\lbrace\sum_{k=0}^{K-1}(\vec{x}_k^i)^T\mat{Q}_x^i\vec{x}_k^i+(\vec{u}_k^i)^T\mat{Q}_u^i\vec{u}_k^i\right\rbrace,
\end{equation}
where $\mat{Q}_x^i$ and $\mat{Q}_u^i$ are positive semi-definite and positive definite matrices of appropriate dimensions, respectively.

The sensor-controller link is closed over the communication network. The control unit of each sub-system $i\in\mathset{L}$ includes a controller $\mathset{C}^i$ that generates, at a time $k$, the control signal~$\vec{u}_k^i$, and an estimator $\mathset{E}^i$ that computes state estimation~$\hat{\vec{x}}_k^i$ in case $\vec{x}_k^i$ is not accessible. We assume that sensors measure perfect copies of their corresponding sub-system's state information. %, i.e. no measurement noise exists and the output matrices are all identities. 
The control input $\vec{u}_k^i$ is generated according to a control law $\xi^i\!=\!\lbrace \xi_k^i\rbrace$, described by causal mappings $\xi_k^i$ from the $i^{\text{th}}$-loop observation history at time $k$ to the respective control input. We allow the control law to depend on the complete observation history of the received information at the control side. In addition, we do not restrict the control inputs to remain constant in between successful transmissions. The combination of local control laws is denoted by $\vec{\xi}=\{\xi^1,\ldots,\xi^L\}\in\Xi$, where $\Xi$ denotes the set of all admissible control laws.

Each sensor has a sampler attached to decide when to transmit the current state information to the controller. The sampling decision is denoted by $\delta_k^i\in\lbrace 0,1\rbrace$, where $\delta_k^i\!=\!1$ indicates that $\vec{x}_k^i$ is transmitted, and $\delta_k^i\!=\!0$ indicates otherwise. The transmission induces a network packet of $r_k^i$ information units\footnote{Information units can, e.g., be bits, Bytes or \glspl{SDU}.}, which is forwarded to the network for transport to the controller\footnote{We generally assume time-varying $r_k^i$ may differ among loops. In many cases we can make the valid assumption that $r_k^i = r^i$ $\forall k$, or $r_k^i = r$ $\forall i,k$.}. The sampling decision is the outcome of a sampling law $\varphi^i$, which maps the history of state observations into $\delta_k^i$. Sampling laws are all aggregated~in~a~vector $\vec{\varphi}\!=\!\{\varphi^1\!,\ldots,\varphi^L\}\!\in\!\Phi$, and $\Phi$ is the set of all admissible laws.
\vspace{-6.8mm} 
\subsection{Network Model}
The network is a set of nodes $\mathset{N}\!=\!\{1,2,\ldots\}$, each representing a transmitting or receiving device. Each sensor is attached to a source node $s_i\!\in\!\mathset{N}$ and each controller to a target node $t_i\!\in\!\mathset{N}$. In general, there might be loops that share a source or target node, such that we define the set $\mathset{U}_n \!=\!\lbrace i\in\mathset{L}\!: s_i \!=\! n\rbrace$ of loops that have a specific node~$n$ as their source node. Further, there might be nodes that are neither source, nor target but can forward data. Each loop is associated with a dedicated path $\mathset{Z}_i\!=\!\lbrace (s_i,n_{i_1}),...,(n_{i_{l-1}},t_i)\rbrace$, i.e., a sequence of links $(n_{i_{d-1}},n_{i_d})\in\mathcal{N}\times\mathcal{N}$ connecting the source and target nodes, over which its corresponding data is transported. The path is kept fixed and determined a priori, therefore routing is not part of the given problem.

The nodes are coupled by a link state matrix $\mat{Q}\in\mathset{Q}$, where each element $Q_{mn}$ denotes the link state for transmissions from node $m$ to $n$ and $\mathset{Q}$ is the set of states that $\mat{Q}$ may assume. For example, in a wireless communication system with single antennas $Q_{mn}\in\lbrack 0,1\rbrack$ could be the attenuation coefficient between $m$ and $n$, while $\mat{Q}$ would be the channel matrix taken from the set $\mathset{Q} = \lbrack 0,1\rbrack^{\lvert\mathset{N}\rvert}$ of possible channel matrices. Each node has a set of available actions to choose to transmit data. The combined actions of all nodes are denoted by the matrix $\mat{A}\!\in\!\mathset{A}$, where element $A_{mn}$ denotes the action of link $(m,n)$ and $\mathset{A}$ is the set of all valid matrices. The combination of an action and link state leads to the data transmission. The maximum amount of data that can be transmitted for a defined action is described by the rate function $\mat{R}\!:\! \mathset{Q}\!\times\!\mathset{A}\mapsto\mathbb{R}_+^{\lvert\mathset{N}\rvert\times\lvert\mathset{N}\rvert}$, which is a mapping from the link and action spaces $\mathset{Q}\!\times \!\mathset{A}$ to the node-by-node transmission rate $R_{mn}$. Each element $R_{mn}$ indicates the amount of data that is transmitted from node $m$ to node $n$, expressed in appropriate dimension of information units per slot. We assume no packet losses occur. We discuss this in more detail in Section~\ref{sec:model_justification}.

Assume that the network operates in a time-slotted fashion with slot $\tau$ of width $T_{\tau}$, i.e., $\tau$ refers to the time interval $\left((\tau-1)T_{\tau},\tau T_{\tau}\right]$. We make the assumption that $T_{\tau}\ll T^i$, $\forall i$, i.e., the network operates significantly faster than the control loops. By this assumption, data transportation can be considered to be delay-free from the control loop perspective, although in reality it might require multiple communication slots due to queuing or retransmissions. Again, we discuss the impact and necessity of the real-time assumption in Section~\ref{sec:model_justification}. We assume that the link state $\mat{Q}$ changes from slot to slot according to a stationary process but remains constant in state $\mat{Q}[\tau]$ at slot $\tau$. Further, in each slot a single action choice $\mat{A}$ can be made, which is denoted by $\mat{A}[\tau]$. The combination of action and link state leads to an amount of $\mat{R}[\tau] = \mat{R}(\mat{Q}[\tau],\mat{A}[\tau])$ information units being transmitted among the nodes. We assume the maximum achievable rate is finite for any link, i.e., $\max_{\mathset{A},\mathset{Q}}\lVert\mat{R}(\mat{Q}[\tau],\mat{A}[\tau])\rVert_{\infty}\!<\!\infty$. 

Each node is assumed to have an infinite length transmission buffer whose traffic passes through the node. The buffer is used to store data originating from the corresponding sensor for relaying. We denote the buffer back-log at slot $\tau$, the set of all back-logs for loop $i$, and all back-logs for all loops by $B_n^i[\tau]$, $\mat{B}^i[\tau] = \lbrace B_n^i[\tau]\mathskip\forall n\in\mathset{N}\rbrace$, and $\mat{B}[\tau] \!=\! \lbrace \mat{B}^i[\tau]\mathskip\forall i\!\in\!\mathset{L}\rbrace$, respectively. A source node $s_i$ is assumed to have an infinitely large CC buffer with the back-log denoted by $Y^i[\tau]$. This buffer stores data output of the sampler until it is pushed into the transmission buffer. Technically, the \gls{CC} buffer resides in the transport layer and corresponds to, e.g., the input buffer to a \gls{TCP} socket, whereas the transmission buffer is in the \gls{MAC} layer at the outgoing network interface chip. Transferring data from the \gls{CC} buffer to transmission buffer is done based on a CC mechanism. %, e.g., a \gls{TCP} variant, with the target set to avoid overloading the \gls{MAC} layer.

\begin{remark}
The described network model has been developed in \cite{2006_NOW_backpressure_book} for cross-layer optimization. %A simple system that falls into the model is \gls{TDMA} wireless system with on-off channel model and unit disk-graph interference model. In this, $\mathset{Q} = \lbrace 0,1\rbrace$, where a link state is one if and only if it is available for transmission, i.e., the channel is sufficiently good, and $\mathset{A} = \lbrace 0,1\rbrace$, indicating transmission with fixed power. A choice $A_{mn}[\tau]=1$ then provides a fixed rate $R_{mn}$ to link $(m,n)$ in slot $\tau$ if the channel is available and no interfering link is transmitting concurrently, otherwise, no data is transported. 
By adjusting the link state and action sets, a variety of communication system models can be realized within the sketched model including wireline networks \cite[Ex. 2.1]{2006_NOW_backpressure_book}, wireless network with channel variation and power control \cite[Ex. 2.5]{2006_NOW_backpressure_book}, ad-hoc networks \cite[Ex. 2.6]{2006_NOW_backpressure_book}, satellite downlinks \cite[Ex. 2.4]{2006_NOW_backpressure_book}, and cellular networks with adaptive modulation and coding \cite[Ch. 2.2.3]{2019_Kluegel_Diss}.
\end{remark} 

Define $\mathset{K}^i[\tau] = \lbrace k: (\tau-1)T_\tau < kT^i\leq \tau T_{\tau}\rbrace$ as the set of control time-steps of loop $i$ that fall into a network slot. Then, the amount of input data $r^i[\tau]$ arriving into the \gls{CC} buffer of source node $s_i$ in slot $\tau$ can be expressed as
\begin{equation}
	r^i[\tau] = \sum\nolimits_{k\in\mathset{K}^i[\tau]}\delta_k^ir_k^i.\vspace{-1mm}
\end{equation}
Note that $r^i[\tau]$ can be interpreted as the arrival process to the \gls{CC} buffer of node $s_i$, the exact statistics of which depend on the sampling mechanism $\varphi^i$ and the differences in time scales of the control and communication systems. Having the assumption that $T_\tau$ is much shorter that $T^i$, $r^i[\tau]\!=\!0$ holds in most of the slots. %, whereas this will not be so probable if $T^i \ll T_\tau$. 
In general, what we essentially require is that $r^i=\mathbbm{E}_\tau\left\lbrace r^i[\tau]\right\rbrace$ remains finite. This is not restrictive as $T^i$ and the sampling rates of all control loops are finite. Therefore, $\frac{T^i}{T_{\tau}}$ is non-zero and finite. 
By the Lindley's recursion \cite[Ch. 1]{Asmussen2003}, the evolution of the \gls{CC} buffer back-log becomes
\begin{equation}
	Y^i[\tau] = \left[Y^i[\tau-1] +r^i[\tau]-\mu_{s_i}^i[\tau]\right]^+,
\end{equation}
where $\mu_{s_i}^i[\tau]$ is the admission decision at node $s_i$, i.e., the data amount pushed from $i$'s \gls{CC} buffer into node $s_i$'s transmission queue in slot $\tau$, with the notation $\left[\cdot\right]^+ \triangleq \max\lbrace \cdot,0\rbrace$. 

The admission can be interpreted as service process to the \gls{CC} buffer, with expected service rate $\mu_{s_i}^i = \mathbbm{E}_\tau\left\lbrace \mu_{s_i}^i[\tau]\right\rbrace$. At the same time, the served amount of data acts as arrival to the \gls{MAC} layer transmission buffer. Define the amount of data originating from loop $i$, that has been transmitted over link $(m,n)$ in slot $\tau$, by $R_{mn}^i[\tau]$. Then the following must hold:
\begin{equation}
\sum\nolimits_{i\in\mathset{L}}R_{mn}^i[\tau] \leq R_{mn}[\tau] \mathskip\forall (m,n), \tau.
\end{equation}

Further, for notational consistency, define arrival processes $\mu_n^i[\tau]$ for any combination of $i$, $n$, such that $\mu_n^i[\tau] = 0$ $\forall \tau$ if $n\neq s_i$. Then, the back-log of the transmission buffer for loop $i$ on node $n$, i.e. $B_{n}^i[\tau]$, evolves in time  according to\footnote{For tractability, we assume that data arrivals in a time slot happen just after transmissions, i.e., received data cannot be transmitted in the same slot.}:\vspace{-1.5mm}
\begin{equation}
B_{n}^i[\tau] \!=\!\! \left[\!B_{n}^i[\tau-1] \!+\!\!\sum_{m\in\mathset{N}}\!\!R_{mn}^i[\tau]\!+\!\mu_n^i[\tau]\!-\!\sum_{o\in\mathset{N}}\!R_{no}^i[\tau]\right]^{\!+}\!\!\!.\!\label{eq:mac_queue_evolution}
\end{equation}
Due to the assumed routing on a single path, a node is either a source, in which case the incoming rates $R_{mn}^i[\tau]$ must be zero in all slots, or otherwise $\mu_n^i[\tau]=0$, $\forall \tau$. We further assume that $B_{t_i}[\tau]= 0$ $\forall \tau$, as all data will be forwarded towards the upper layers, i.e., to the controller, without considerable delay. 

Given this model, in each slot the actions $\mat{A}[\tau]$ are chosen according to a scheduling law $\pi\in\Pi$, where  $\pi$ is a mapping from the history of link states $\mathset{H}_{\mat{Q}}[\tau]=\lbrace ...,\mat{Q}[\tau-1],\mat{Q}[\tau]\rbrace$ and queue back-logs $\mathset{H}_{\mat{B}}[\tau]=\lbrace ...,\mat{B}[\tau-1],\mat{B}[\tau]\rbrace$ to an action out of the set $\mathset{A}$, and $\Pi$ is the set of all possible scheduling laws. Similarly, the  admissions $\mu_n^{i}[\tau]$ are chosen according to a \gls{CC} law $\psi^i$, used by loop $i$. All employed \gls{CC} laws are gathered in the vector $\psi\in\Psi$, which is one out of a set of possible laws $\Psi$. Similar to $\pi$, each $\psi^i$ is a mapping from the history of \gls{CC} buffer back-logs $\mathset{H}_{Y^i}[\tau]=\lbrace ...,Y^i[\tau-1],Y^i[\tau]\rbrace$ to an amount of admitted data. Both $\pi$ and $\psi$ can, but do not have to, incorporate the effect of reporting delays by depending on out-dated buffer status and the effect of estimation inaccuracies by including randomness into the decisions. Further, $\pi$ can be realized in a centralized manner or distributively at the nodes.

\vspace{-2mm}

\subsection{Model Justification}\label{sec:model_justification}
Several simplifying assumptions are made in the model, that require further justification. First, all transmissions are assumed to be error-free. As has been comprehensively discussed in \cite[Ch. 2.4.3]{2006_NOW_backpressure_book}, this simplification does not severely limit the results as long as lost data is re-injected into the network by an error-recovery protocol, such as \gls{ARQ}. Re-injection then can be modeled by an equivalent rate reduction of the channel, which falls into the sketched model. 

Second, we assume that end-to-end data transport is finalized within a single control step $T^i$ for all loops. The reason for this assumption is that delays can create a feedback effect that is not yet fully understood. In fact, increased delays can lead to more requests for data transmission by the control loops, which in turn can increase the delay even further due to queuing effects. %The delay assumption corresponds to the timeliness demands of real-time communication, which we hence adopt in this work.
 Nevertheless we expect that our results transfer to the delay-affected case with some additional modifications and find a reasonable performance verified even for the delay-affected case in our simulation results.

The targeted problem remains challenging even with these simplifications. The main challenge arises from the question which control loops should transmit while network can only serve $D\! < \!L$ transmissions per time.
%can be described with an example: Consider the example of $L$ synchronized control loops closed over a network that can only serve $D < L$ transmissions per time step.  
Further, what if network has complex physical and \gls{MAC} layers, randomly varying channels and relays data in multi-hop fashion, such that $D$ is unknown or time-dependent. Other challenges are when control loops are not synchronous and have partial information about each other, or the network is not control-aware. %In all scenarios, how to optimize control performance is the question that we target with the given model.

\vspace{-2mm}

\subsection{Buffer Stability, Capacity and Transport Capacity}
%Here we re-visit several useful notions required for the problem presentation. 
Define $R_{mn}^i \!:= \!\mathbb{E}_\tau\lbrace R_{mn}^i[\tau]\rbrace$ as the time-average of $R_{mn}^i[\tau]$. We write similarly for $r^i$ and $\mu_n^i$. Let the buffers in the network be modeled as queuing systems with the input data as arrival process and the outgoing data as service process. From queuing theory, we take the definition of \textit{queue stability} \cite{2006_NOW_backpressure_book,2010_Neely_StochNetOpt}, for a queue with back-log $B_n^i[\tau]$, which is defined by
\begin{equation}
	\limsup_{N\to\infty}\frac{1}{N}\mathbbm{E}\left\lbrace\sum\nolimits_{\tau=1}^{N}B_n^i[\tau]\right\rbrace < \infty.
\end{equation}
If the expectations of the arrival and departure processes are $r^i$ and $\mu_{s_i}^i$, respectively, then under some loose  \textit{admissibility} assumptions \cite[Def.'s 3.4-3.5]{2006_NOW_backpressure_book}, $Y^i[\tau]$ is stable iff $r^i < \mu_{s_i}^i$ \cite[Lemma 3.6]{2006_NOW_backpressure_book}. This notion can be extended to a network, which we call \textit{stable} if all network queues are stable \cite[Def. 3.2]{2006_NOW_backpressure_book}. A network is stable in the queuing sense %(all incoming data rates are on average serviced by an appropriate outgoing data rate),
 iff \cite{2010_Neely_StochNetOpt}
\begin{equation}\label{eq:queue-stability-condition}
	\mu_{n}^i + \sum\nolimits_{m\in\mathset{N}}R_{mn}^i < \sum\nolimits_{o\in\mathset{N}}R_{no}^i\mathskip \forall i,n.
\end{equation}
If a network is unstable then a node with a bottleneck link exists. %that receives more data than its capacity.  
Assume that we need to operate the network to meet average per-link rate targets $R_{mn}$, $\forall m,n$. Then we can define the feasible region for all targets, i.e., the \textit{Network Capacity} $\mathset{C}$, \cite{2006_Lin_Schroff_Backpressure_Scheduling}. Let the actions $\mat{A}[\tau]$ in each slot be chosen according to a scheduling law $\pi\in\Pi$. Then from \cite{2006_NOW_backpressure_book}, $\mathset{C}$ is defined as
\begin{eqnarray*}
	\mathset{C} &=&\left\lbrace \mat{R}:  R_{mn}\geq 0, \mathskip R_{nn} = 0  \mathskip \forall m,n;\right.\\
	&&\left.\exists\pi\in\Pi: R_{mn}\leq \mathbbm{E}_\tau\lbrace R_{mn}(\mat{Q}[\tau],\mat{A}[\tau])\rbrace\mathskip\forall m,n\right\rbrace,\notag
\end{eqnarray*}
and is the set of expected per-link rates that can be provided by at least one scheduling law. The first row ensures that all rates are non-negative and there is no self-communication. An important property is that $\mathset{C}$ is a convex set \cite{2006_NOW_backpressure_book}. 

Here we take a transport-layer perspective, in which the exact routes and \gls{MAC} layer procedures are not of interest. Given that all paths $\mathset{Z}_i$ are fixed, we extend the MAC layer network capacity towards a transport-layer capacity; the \textit{Transport Capacity} $\Lambda$. Define the set of valid per-loop link rates $\mathset{R}_i$ as $\mathset{R}_i = \left\lbrace \mat{R}^i: R_{mn}^i\geq 0;\mathskip R_{mn}^i = 0 \mathskip\forall (m,n)\notin \mathset{Z}_i\right\rbrace.$
Considering that $\mu_n^i$ is an expected end-to-end data rate, the transport capacity is defined as in \cite{2006_NOW_backpressure_book}:
\begin{subequations}\label{eq:transport_capacity}
\begin{eqnarray}
\Lambda &=& \left\lbrace \vec{\mu}\geq\vec{0} : \mu_n^i = 0\mathskip\forall n \neq s_i,\right.\label{eq:transport_validity}\\
		  && \exists \mat{R}\in\mathset{C}\text{ and }\mat{R}^i\in\mathset{R}_i,\label{eq:route_validity}\\
		\text{ s.t. } \hspace{-8mm}&& \sum\nolimits_{i\in\mathset{L}}R_{mn}^i \leq R_{mn}\mathskip\forall (m,n)\label{eq:mac_fesasibility}\\
		  && \left.\mu_n^i + \sum\nolimits_{m\in\mathset{N}}R_{mn}^i \leq \sum\nolimits_{o\in\mathset{N}}R_{no}^i\right\rbrace,\label{eq:queue_stability_constr}
\end{eqnarray}
\end{subequations}
where $\vec{0}$ is a zero vector of compatible dimension and $\geq$ denotes elementwise comparison. Vector $\vec{\mu}$ contains the served, expected end-to-end rate for each communication flow $i$. The constraint \eqref{eq:transport_validity} ensures that rates are only provided to flows that originate at the respective node, while \eqref{eq:route_validity} demands that all \gls{MAC} layer average rates are chosen from the network capacity and hence achievable by a scheduling law, as well as that the per-link rate assignments comply with the chosen routes. The third constraint \eqref{eq:mac_fesasibility} ensures that the loop rates comply with the link rates, and \eqref{eq:queue_stability_constr} ensures network stability\footnote{\label{footnote:note2} Network stability condition is formally introduced in (\ref{eq:queue-stability-condition}). Comparing it with the constraint \eqref{eq:queue_stability_constr}, ``$\leq$'' is an approximation of ``$<$'' to ease the development. Technically, we would have to add an arbitrarily small $\varepsilon > 0$ to the left hand side and take the limit $\varepsilon \to 0$, which leads to the same results.}. 
Note that $\mathset{C}$ and $\mathset{R}_i$ are convex, and so are the constraints \eqref{eq:transport_validity}--\eqref{eq:queue_stability_constr}, hence $\Lambda$ is convex as it is formed by the intersection of convex sets.

Given the transport capacity, we can demand that all loop rates $r^i$ should be served by the network by simply ensuring%\footnote{\label{footnote:note2} The ``$\leq$'' here is an approximation of ``$<$'' to ease development. Technically, we would have to add an arbitrarily small $\varepsilon > 0$ to the left hand side and take the limit of $\varepsilon \to 0$, which however leads to the same results.}%
\begin{equation}
	r^i \leq \mu_{s_i}^i\mathskip\forall i; \bigmathskip \mat{\mu}\in\Lambda.\label{eq:abstractly_bounded_rates}
\end{equation}
This formulation %comprises a transport-layer perspective and helps to
 abstracts away the technical complexity of including the underlying network and \gls{MAC} layers explicitly. %, specifying the routes and provided data rates per slot.

\section{Problem Statement \& Assessment}
\label{sec:problemstatement}
We now construct our target optimization problem, so-called global optimization problem (GOP). The goal is to maximize control performance in form of a weighted cost making use of both control and network parameters. It is formulated as
\begin{equation}\label{first-GOP}
\!\!\text{GOP:} \min\limits_{\varphi,\xi,\psi,\pi} \sum_{i\in\mathset{L}} w^iJ^i,\text{ s.t. }\varphi\in\Phi,\xi\in\Xi,\psi\in\Psi,\pi\in\Pi.\!\!\!\!
\end{equation}
GOP minimizes the weighted sum of local \gls{LQG} costs, over all possible sampling, control, CC and scheduling laws.

\vspace{-2mm}

\subsection{Decomposition of Control and Networking}
The GOP considers joint optimization over networking and control parameters. As this constitutes an impractical solution from the global perspective, we aim at finding ways to separate different aspects from one another. To do so, we interpret $J^i = J^i(\varphi,\xi,\psi,\pi)$ as set-function of the considered policies. Then, we can use the general equality \cite{2004_Boyd}
\begin{eqnarray}
	\!\!\!\min\limits_{\varphi,\xi,\psi,\pi} \sum\nolimits_{i\in\mathset{L}} w^iJ^i &\!\!\!=\!\!\!&  \min\limits_{\psi,\pi}v(\psi,\pi), \label{eq:master_problem}\\
	\!\!\!\text{where }v(\psi,\pi)&\!\!\!=\!\!\!&\inf\limits_{\Phi\times\Xi} \sum\nolimits_{i\in\mathset{L}} w^iJ^i(\varphi,\xi,\psi,\pi). \label{eq:primal_problem}
\end{eqnarray}
That is, for fixed CC mechanism $\psi$ and scheduling law $\pi$, the cost functions are optimized over the possible sampling and control laws $\varphi$ and $\xi$, respectively. The mapping of $\{\psi$, $\pi\}$ to its minimum weighted sum cost, i.e., given the optimal $\varphi$ and $\xi$, can be interpreted as set-function $v(\psi,\pi)$ over $\Psi\times\Pi$, and the network parameters are optimized in the outer loop.

The problem \eqref{eq:master_problem} is denoted the \textit{master problem}, and \eqref{eq:primal_problem} the \textit{primal problem}. Note that technically, we would have to restrict $\{\psi$,$\pi\}$ to a set that renders the primal problem feasible. However, as it is an unconstrained problem, the feasible sets are simply the introduced sets $\Psi$ and $\Pi$, respectively. 

We can bring this abstract formulation into a more technical form that resides on transport layer. For this, we leverage results that have been derived in \cite{Molin2014}, where the problem 
\begin{equation}
	\min\limits_{\varphi,\xi}\sum\nolimits_{i\in\mathset{L}} J^i \text{ s.t. }\sum\nolimits_{i\in\mathset{L}} r^i\leq c\label{eq:control_problem_molin}
\end{equation}
is considered. It is argued that the average rate $r^i(\varphi,\xi)$ and cost $J^i(\varphi,\xi)$ depend on the chosen pair $\{\varphi$, $\xi\}$. 	Hence the region of all feasible tuples $(J^i,r^i)$ can be constructed as
\begin{equation}
	\mathset{J}_i = \left\lbrace (J_{\ast}^i,r_{\ast}^i): \exists \varphi,\xi\text{ s.t. }J^i(\varphi,\xi) = J_{\ast}^i; r^i(\varphi,\xi)=r_{\ast}^i\right\rbrace.\notag
\end{equation}
Note that $\mathset{J}_i$ is shown to be convex \cite{Molin2014}. Then, the Pareto curve 
\begin{equation}\
	J^i(r^i) = \inf\lbrace J_{\ast}^i:  (J_{\ast}^i,r^i)\in\mathset{J}_i\rbrace
\end{equation}
represents a convex function taking the rate as input to~determine the optimal cost. The curve $J^i(r^i)$ then corresponds to the cost under an optimal choice $\varphi$, $\xi$ such that $r^i(\varphi,\xi)\!=\!r_{\ast}^i$.

Adopting this knowledge, we can further re-state the master problem by referring to the feasible region of end-to-end rates that can be served by a combination $(\psi,\pi)$ of CC and scheduling. This region is exactly the transport capacity $\Lambda$. Then, the master problem simply reduces to
\begin{equation}
	\min\limits_{\vec{\mu},\vec{r}} \sum\nolimits_{i\in\mathset{L}} w^iJ^i(r^i)\text{ s.t. }r^i\leq\mu_{s_i}^i\mathskip\forall i;\mathskip \vec{\mu}\in\Lambda.\label{eq:gop_in_rates}
\end{equation}
Note that, technically, in problem (\ref{eq:gop_in_rates}), we do not directly optimize over the networking and control policies, but rather over a set of feasible end-to-end rates and achievable cost values for which such policies exist. However, we can deduce the optimal policies from the argument minimizing the problem.

\vspace{-2mm}

\subsection{Optimal Control}
\label{subsec:OptimalControl}
Due to the similarity of \eqref{eq:control_problem_molin} and \eqref{eq:gop_in_rates}, the results of \cite{Molin2014} are directly transferable to solve the primal problem. As effect of the sum-structure in the objective, for given fixed $\mu_{s_i}^i$, the primal problem can be decomposed into per-loop problems as
\begin{equation}\label{GOP:control-structure}
\min\limits_{\varphi^i,\xi^i}w^i J^i(r^i)\text{ s.t. } r^i\leq \mu_{s_i}^i.
\end{equation}

	Inspired by \cite{Molin2014}, we study the structural properties of the optimal control and event-triggered sampling policy. In fact, \eqref{GOP:control-structure} can be solved by scalarization approach using a fixed Lagrange multiplier $\lambda_i\geq 0$, yielding the unconstrained problem
	\begin{equation}
	\min\limits_{\varphi^i,\xi^i}w^i J^i(r^i) + \lambda_i\left(r^i-\mu_{s_i}^i\right)\hat{=}\min\limits_{\varphi^i,\xi^i}w^i J^i(r^i) + \lambda_ir^i,\notag
	\end{equation}
	where $\lambda_i$ is referred to as ``communication price''. For fixed $\lambda_i$, the optimal control policy is derived as a certainty equivalence controller in combination with a model-based estimator. For time-varying $\lambda_i$, an adaptation model is proposed in \cite{Molin2014} that proposes an adaptive pricing model by use of a gradient ascent on the dual problem. The results of this article are extendable to the adaptive pricing following the ideas of \cite{Molin2014}.
	
	It is observed from the relaxed problem \eqref{eq:gop_in_rates} that the optimal policies $\xi^i$ and $\varphi^i$ are solely characterized by the transmission rate $r^i$ and the cost $J^i$ of sub-system $i$. Hence, we will search for the feasible region of the pairs $(J^i,r^i)$ with respect to the dominating class of control and sampling strategies to form the Pareto optimal policies. Having the class of dominating strategies found, we can then confine our search within this class of narrowed down admissible strategies to solve the optimization problem \eqref{eq:gop_in_rates} without loosing optimality.
	
	It is discussed in \cite{molin2013optimality} that any pair of policies $(\xi^i,\varphi^i)$ with the certainty equivalence controller is dominating. Hence, the optimal control policy can be expressed as the causal mapping of the observation history stored at the controller side, i.e.,
	\begin{equation}\label{eq:contro-policy-CE}
	\vec{u}_k^i=\xi_k^{i,\ast}(Z_k^i)=-\vec{K}^i_\ast \;\mathbb{E}[\vec{x}_k^i|Z_k^i],
	\end{equation}
	where $Z_k^i$ denotes the observation history at the control side from the initial time until time $k$, and $\mathbb{E}[\vec{x}_k^i|Z_k^i]$ denotes the optimal state estimation at time $k$ at control side given $Z_k^i$. %Note that the optimal control policy can be characterized as in expression \eqref{eq:contro-policy-CE} under the partially nested information assumption, i.e, the set of information $Z_k^i$, accessible for the control unit of loop $i$ at any time-step $k$, is essentially a subset of the information available for the sampling unit of the same loop at that time-step. 
	The optimal control gain $\vec{K}^i_\ast$ can then be computed as follows
	\begin{equation}
	\vec{K}^i_\ast= (\vec{B}^{i^\textsf{T}}\vec{P}^i\vec{B}^i+\vec{Q}_u^i)^{-1}\vec{B}^{i^\textsf{T}}\vec{P}^i\vec{A}^i,\label{eq:opt_control_gain}
	\end{equation}
	where $\vec{P}^i$ solves the succeeding algebraic Riccati equation
	\begin{align*}
	&\vec{P}^i\!=\!\vec{Q}_x^i\!+\!\vec{A}^{i^\top}\!\left(\vec{P}^i\!-\vec{P}^i\vec{B}^i\!(\vec{Q}_u^i+\vec{B}^{i^\top} \!\vec{P}^i\vec{B}^i)^{-1}\vec{B}^{i^\top} \vec{P}^i\right)\!\vec{A}^i.
	\end{align*}

	We denote the binary variable $\gamma_k^i$ as the \textit{delivery indicator}, i.e., if $\delta_k^i\!=\!1$, then $\gamma_k^i\!=\!1$ indicates that $\vec{x}_k^i$ is successfully delivered at time-step $k$, and otherwise if $\gamma_k^i\!=\!0$. Clearly, $\gamma_k^i=0$ if $\delta_k^i=0$. The optimal state estimation, given the information set $Z_k^i$ and $\mathbb{E}[\vec{x}_{0}^i]$, can then be expressed as
	\begin{equation}\label{eq:estimation_process}
	\mathbb{E}[\vec{x}_k^i|Z_k^i]=\begin{cases} \vec{x}_k^i, & \quad\delta_k^i \gamma_k^i=1,\\
	\mathbb{E}[\vec{x}_k^i|Z_{k-1}^i], & \quad\text{otherwise},
	\end{cases}
	\end{equation}
	where, from \eqref{eq:loop_dynamics} and \eqref{eq:contro-policy-CE}, we have
	\begin{align}
	\mathbb{E}[\vec{x}_k^i|Z_{k-1}^i]&\!=\!\mathbb{E}[\vec{A}^i\vec{x}_{k-1}^i-\vec{B}^i\vec{K}^i_{\ast}\mathbb{E}[\vec{x}_{k-1}^i|Z_{k-1}^i]\!+\!\vec{w}_k^i|Z_{k-1}^i]\notag\\
	&\!=\!(\vec{A}^i-\vec{B}^i\vec{K}^i_{\ast})\;\mathbb{E}[\vec{x}_{k-1}^i|Z_{k-1}^i].\label{eq:opt_estimator}
	\end{align}
The variable $\gamma_k^i$ in (\ref{eq:estimation_process}) represents the effects of network
actions on transmitted data packets. Assuming that the communication network has much finer slotted periods compared to the control sampling periods, and also assuming that there are no (net) packet losses, it holds that $\gamma_k^i = \delta_k^i$, i.e., any packet is delivered within the corresponding control sampling slot if it is transmitted, and hence $\delta_k^i\gamma_k^i = \delta_k^i$ in \eqref{eq:estimation_process}. In fact, we assume that the time duration required for the entire data transportation including contention resolution, queuing, retransmissions, and node allocation, is negligible from the control perspective. This assumption remains valid for a wide range of NCS applications wherein the dynamics of the processes are not super fast, e.g. smart homes, traffic control, electric power flow, district heating systems and typical chemical process control. In these applications, the required sampling rates of the dynamical processes are typically $10$Hz or lower, while today's communication technology can guarantee much faster data handling and transmission with negligible delay ($<\!1$ms). This idealization leads to expressing the estimation process as in (\ref{eq:estimation_process}) where $\delta_k^i\gamma_k^i$ is being substituted by $\delta_k^i$.
%	\begin{equation}\label{eq:estimation_process_2}
%	\mathbb{E}[\vec{x}_k^i|Z_k^i]=\begin{cases} \vec{x}_k^i, & \quad\delta_k^i=1,\\
%	\mathbb{E}[\vec{x}_k^i|Z_{k-1}^i], & \quad\text{otherwise}.
%	\end{cases}
%	\end{equation}
We define $\vec{e}_{k|k-1}^i=\vec{x}_k^i-\mathbb{E}[\vec{x}_k^i|Z_{k-1}^i]$ as the estimation error computed at the sampling unit of sub-system $i$ at time-step $k$ given the information $Z_{k-1}^i$. Note that $\vec{e}_{k|k-1}^i$ is computed before the sampler decides on $\delta_k^i$. Therefore, using \eqref{eq:loop_dynamics}, \eqref{eq:estimation_process} and $\delta_k^i\gamma_k^i\! =\! \delta_k^i$ and knowing that $\delta_k^i\!=0$ results in $\mathbb{E}[\vec{x_k}^i|Z_k^i]\!=\!\mathbb{E}[\vec{x}_k^i|Z_{k-1}^i]$, the dynamics of the one-step ahead estimation error becomes
	\begin{align}%\nonumber
	\vec{e}_{k+1|k}^i&=\left(1-\delta_k^i\right)\vec{A}^i \vec{e}_{k|k-1}^i +\vec{w}_k^i.\label{eq:estimario_dynamics}
%	\vec{e}_{k+1|k}^i&=\vec{x}_{k+1}^i-\mathbb{E}[\vec{x}_{k+1}^i|Z_{k}^i]=\left(1-\delta_k^i\right)\vec{A}^i \vec{e}_{k|k-1}^i +\vec{w}_k^i.\label{eq:estimario_dynamics}
%	&=\vec{A}^i\vec{x}_{k}^i-\vec{B}^i\vec{K}^i_{\ast}\mathbb{E}[\vec{x}_k^i|Z_k^i]+\vec{w}_k^i\\ \nonumber
%	&- \vec{A}^i\mathbb{E}[\vec{x}_{k}^i|Z_{k}^i] +\vec{B}^i\vec{K}^i_{\ast}\mathbb{E}[\mathbb{E}[\vec{x}_k^i|Z_k^i]|Z_{k}^i]-\mathbb{E}[\vec{w}_k^i|Z_k^i]\\\nonumber
%	&=\vec{A}^i\left(\vec{x}_k^i-\mathbb{E}[\vec{x}_{k}^i|Z_{k}^i]\right)+\vec{w}_k^i\\
%	&=\left(1-\delta_k^i\right)\vec{A}^i \vec{e}_{k|k-1}^i +\vec{w}_k^i,
	\end{align} 

\begin{remark}
To derive the optimal sampling policy we use the results presented in \cite{Molin2014,6475979} and assume that for the event-based remote estimation the optimal event-triggered sampling law is symmetric, if: 1) the noise distribution is zero-mean, unimodal and symmetric, and, 2) distribution of $\vec{x}^i_0$ is symmetric around $\mathbb{E}[\vec{x}^i_0]$, $\forall i$. Despite some efforts in, e.g. \cite{6487384,Rabi:2012} that hint the optimality of symmetric event-triggered sampling laws for higher dimensions, this result is formally proved only for first-order LTI systems \cite{5744106}. However, it is discussed in~\cite{5717890} that even if the optimal event-triggered sampling law is not symmetric for higher order systems, then an extra bias term will be added to the optimal estimator which has no compromising effect on derivation of our following results. %The main advantage of the event-triggered sampling law being symmetric is that the optimal estimator becomes identical to that of optimal time-triggered estimator. This means the estimator does not receive any information in between event-triggered transmissions and hence the optimal estimation can be computed according to the linear prediction of the form $\mathbb{E}[\vec{x}_k^i|\vec{x}_{\bar{k}}^i]$ where $x_{\bar{k}}^i$ is the latest received state update associated to sub-system $i$, and $\bar{k}\leq k$.

\end{remark}	
	
Employing \eqref{eq:contro-policy-CE}, \eqref{eq:estimation_process} and $\delta_k^i\gamma_k^i = \delta_k^i$, the local LQG cost function $J^i$ in \eqref{eq:individual_cost} can be equivalently expressed as follows:
	\begin{equation*}
	J^i\!=\!\textsf{Tr}\!\left(\vec{P}^i\vec{Z}^i\right)+\limsup_{K\to\infty}\frac{1}{K}\mathbbm{E}\!\left[\!\sum_{k=0}^{K-1}\!\left(1-\delta_k^i\right)\vec{e}_{k|k-1}^{i^T}\mat{Q}_e^i\vec{e}_{k|k-1}^i\right]
	\end{equation*}
	where $\vec{Q}_e^i\!=\!\vec{K}^{i^\top}_{\ast}\!\left(\vec{Q}_u^i+\vec{B}^{i^\top}\!\vec{P}^i\vec{B}^i\right)\!\vec{K}^i_{\ast}$, and $\vec{Z}^i$ is the covariance matrix of the corresponding noise process with realization $\vec{w}_k^i$. The optimal sampling law $\varphi^{i,\ast}_k$ can then be computed as
	\begin{equation}\label{eq:optimal_sampling}
	\varphi^{i,\ast}_k (e_{k|k-1}^i)= \text{arg}\min_{\varphi^i\in \Phi^i} J^i(e_{k|k-1}^i),
	\end{equation}
	where $\Phi^i$ denotes the class of all admissible local sampling policies. According to the discussions in \cite{Molin2014}, (\ref{eq:optimal_sampling}) can be solved via scalarization method \cite{Hernandez-Lerma2004} using a fixed Lagrange multiplier $\lambda_i\geq 0$, yielding an unconstrained problem. Hence, the Pareto curve of the pair $(J^i,r^i)$ can be characterized as
	\begin{equation}\label{eq:pareto} 
	\min\nolimits_{\varphi^{i,\ast}_k} \{J^i+\lambda_ir^i\}.
	\end{equation}
	%with $\lambda_i$ the Lagrange multiplier that can be interpreted as the cost of communication. 
	To solve the problem (\ref{eq:pareto}), we need to characterize the triggering law $\phi_k^i$. From (\ref{eq:loop_dynamics}), (\ref{eq:contro-policy-CE}) and (\ref{eq:estimario_dynamics}) one can simply derive the closed-loop dynamics of a sub-system $i$ as follows:
\begin{equation}\label{eq:closed-loop}
\vec{x}_{k+1}^i = (\mat{A}^i- \mat{B}^i\mat{K}_{\ast}^i)\vec{x}_{k}^i+(1-\delta_k^i)\mat{B}^i\mat{K}_{\ast}^i \vec{e}_{k|k-1}^i + \vec{w}_k^i.
\end{equation}
According to (\ref{eq:closed-loop}), having the stabilizing gain $\vec{u}_k^i$, the process state $\mat{K}_{\ast}^i$ is stable and controlled by the estimation error state $\vec{e}_{k|k-1}^i$. Hence, to design the triggering law, we confine our search to error-dependent policies. To introduce the class of admissible triggering laws $\Phi^i$, assume an arbitrary variable $M^i(\lambda_i)\!\in \!\mathbb{R}^+$, such that $\varphi^{i,\ast}_k (\vec{e}_{k|k-1}^i)\!=\!0$ for $\|\vec{e}_{k|k-1}^i\|_2\!\leq\! M^i(\lambda_i)$, and $\varphi^{i,\ast}_k (\vec{e}_{k|k-1}^i)\!=\!1$ for $\|\vec{e}_{k|k-1}^i\|_2\!>\! M^i(\lambda_i)$. Since $M^i(\lambda_i)$ can be selected freely, i.e. very small or very large, it does not impose any additional constraint on the class of admissible scheduling policies. Under this assumption, the problem (\ref{eq:pareto}) can be solved by duality using value iteration \cite{Hernandez-Lerma2004,Hernandez-Lerma:2001:AMC:516568}. Moreover, according to \cite{Molin2014}, the optimal event-triggered sampling law $\varphi^{i,\ast}_k (\vec{e_}{k|k-1}^i): \mathbb{R}^+\cup \{0\}\mapsto \{0,1\}$ is a stationary mapping from the estimation error to a transmission decision. Finally, the optimal sampling of sub-system $i$ at time-step $k$ can be expressed as $\delta_k^i = \mathbbm{1}\lbrace  \|\vec{e}_{k|k-1}^i\|_2 > M^i(\lambda_i)\rbrace$.
\vspace{-2mm}

\subsection{Dual Decomposition}
\label{subsec:DualDecomposition}
We now target network optimization using the formulation in \eqref{eq:gop_in_rates}. The employed approach is a dual decomposition with on-line adaptation of the Lagrange multipliers. Remind that both $J^i(r^i)$ and $\Lambda$ are convex, hence so is the problem \eqref{eq:gop_in_rates}. Further, Slaters' constraint qualification \cite[Ch. 5.2.3]{2004_Boyd} is guaranteed to hold for any given positive rate. This results in zero duality gap and strong duality then holds. By explicitly formulating the constraints that define $\Lambda$, the GOP becomes:
\begin{subequations}
\begin{eqnarray}
	&\min\limits_{\vec{\mu},\vec{r}} \sum_{i\in\mathset{L}} w^iJ^i(r^i)&\label{eq:gop_expl_util}\\
	\text{s.t.}&r^i\leq \mu_{s_i}^i;&\label{eq:gop_expl_cc1}\\
	&\mathskip r^i,\mu_{s_i}^i\geq 0\mathskip\forall i;\mathskip \mu_n^i=0\mathskip\forall n\neq s_i;&\label{eq:gop_expl_cc2}\\
	&-------------&\notag\\
	&\mu_n^i + \sum_{m\in\mathset{N}}R_{mn}^i \leq \sum_{o\in\mathset{N}}R_{no}^i\mathskip \forall n,i;&\label{eq:gop_expl_stability}\\
	& \sum_{i\in\mathset{L}}R_{mn}^i \leq R_{mn}\mathskip\forall (m,n);&\label{eq:gop_expl_rate_compliance}\\
	& \mat{R}\in\mathset{C},\mathskip \mat{R}^i\in\mathset{R}_i\mathskip\forall i.&\label{eq:gop_expl_remaining}
\end{eqnarray}
\end{subequations}
As indicated by the dashed line, the functionality of the problem covers two networking layers. The optimizing function \eqref{eq:gop_expl_util} and the first two constraints \eqref{eq:gop_expl_cc1}, \eqref{eq:gop_expl_cc2} reside in the transport layer, because the impacted variables are the rates injected by the control sampler and the \gls{CC} admission rates $\mu_{s_i}^i$. Constraints \eqref{eq:gop_expl_stability}--\eqref{eq:gop_expl_remaining} reside on the \gls{MAC} layer, as they constrain the per-link rates. When explicitly relaxing only the constraints \eqref{eq:gop_expl_cc1} and \eqref{eq:gop_expl_stability}, the Lagrangian function becomes
\begin{eqnarray}
	L(\vec{r},\vec{\mu},\mat{R}^i,\vec{\lambda},\vec{q}) &=& \sum\nolimits_{i\in\mathset{L}}\left[ w^iJ^i(r^i) + \lambda_i(r^i-\mu_{s_i}^i)\right]\notag\\
	&&\hspace{-2.5cm}+\sum\limits_{n\in\mathset{N}}\sum_{i\in\mathset{L}}q_n^i\left(\mu_n^i + \sum_{m\in\mathset{N}}R_{mn}^i - \sum_{o\in\mathset{N}}R_{no}^i\right).\notag
\end{eqnarray}
Variables $\lambda_i$ and $q_n^i$ are non-negative Lagrangian multipliers, respectively. We can now construct the \gls{KKT} condition \cite[Ch. 5.5.3]{2004_Boyd} for $\mu_n^i$ by demanding $\partial L /\partial \mu_n^i = 0$, which yields that $\lambda_i = q_{s_i}^i$. Using this, both, $\vec{\mu}$ and $\vec{\lambda}$ can be eliminated from the Lagrangian, which now is:
\begin{eqnarray}
	L(\vec{r},\mat{R}^i,\vec{q}) &=& \sum\nolimits_{i\in\mathset{L}} \left[w^iJ^i(r^i) + q_{s_i}^ir^i\right]\label{eq:final_lagrange}\\
	& +&\sum\limits_{n\in\mathset{N}}\sum_{i\in\mathset{L}}q_n^i\left(\sum_{m\in\mathset{N}}R_{mn}^i - \sum_{o\in\mathset{N}}R_{no}^i\right).\notag
\end{eqnarray}
Let $\Omega = \left\lbrace \vec{r},\mat{R}^i: \exists\mat{R}\text{ s.t. }\text{\eqref{eq:gop_expl_cc2}, \eqref{eq:gop_expl_rate_compliance}, \eqref{eq:gop_expl_remaining} hold}\right\rbrace$ be the domain of $L(\vec{r},\mat{R}^i,\vec{q})$ over the primal variables $\vec{r}$ and $\mat{R}^i$. 
%\begin{equation}
%	\Omega = \left\lbrace \vec{r},\mat{R}^i: \exists\mat{R}\text{ s.t. }\text{\eqref{eq:gop_expl_cc2}, \eqref{eq:gop_expl_rate_compliance}, \eqref{eq:gop_expl_remaining} hold}\right\rbrace.
%\end{equation}
Then, the dual function reduces to
\begin{equation}
	\Theta(\vec{q}) = \inf\limits_{\left(\vec{r},\mat{R}^i\right)\in\Omega}L(\vec{r},\mat{R}^i,\vec{q}).\notag
\end{equation}
Due to strong duality, the GOP can be solved in the dual domain, i.e., by solving $\max_{\vec{q}\geq \vec{0}}\Theta(\vec{q})$, which is a convex problem. While we cannot provide a closed-form expression for the dual function, we can evaluate its value for given $\vec{q}$ by solving a minimization problem on $L(\vec{r},\mat{R}^i,\vec{q})$. For fixed multipliers $\vec{q}$, the Lagrangian \eqref{eq:final_lagrange} has two additive parts that depend on different variables $\vec{r}$ and $\mat{R}^i$. Thus, both parts can be optimized independently of each other. In $\vec{r}$, the optimization problem decomposes into independent problems of the form:
\begin{equation}
	\min\limits_{r^i\geq 0}w^iJ^i(r^i) + q_{s_i}^i r^i,\label{eq:control_problem}
\end{equation}
which is an unconstrained problem for a single loop. The solution for (\ref{eq:control_problem}) is investigated in \cite{Molin2014} and is discussed above. 

By re-ordering the second sum of $L(\vec{r},\mat{R}^i,\vec{q})$, converting the minimization into a maximization and explicitly stating the constraints of $\Omega$, the optimization problem in $\mat{R}^i$ becomes:
\begin{eqnarray}
	&\max\limits_{\mat{R}^i,\mat{R}} \sum\limits_{m\in\mathset{N}}\sum\limits_{n\in\mathset{N}}\sum\limits_{i\in\mathset{L}}R_{mn}^i\left(q_m^i - q_n^i\right)&\label{eq:MAC_problem}\\
	\text{ s.t. }&\sum\limits_{i\in\mathset{L}}R_{mn}^i \leq R_{mn}\mathskip\forall (m,n);&\\
	&\mat{R}\in\mathset{C},\mathskip \mat{R}^i\in\mathset{R}_i\mathskip\forall i.&
\end{eqnarray}
Parts of this problem can be solved analytically. Note that the majority of $R_{mn}^i$ are zero due to the constraint set $\mathset{R}_i$, which constrains communication to the routed path of each loop. For fixed $R_{mn}$, the problem decomposes into independent problems for each link $(m,n)$. In particular, a weighted sum of those per-loop rates $R_{mn}^i$ that are allowed to use link $(m,n)$ needs to be maximized. Clearly, if $q_m^i < q_n^i$, the maximum value is obtained by $R_{mn}^i=0$. Among all those loops $i$ for which $q_m^i \geq q_n^i$, the optimal solution is readily solved as
\begin{equation}
	R_{mn}^i=\begin{cases}R_{mn}, &\text{if }i = \arg\max_i \lbrace q_m^i - q_n^i\rbrace\\ 0, &\text{else}.  \end{cases}
\end{equation}
If several links are tied on the $\arg\max$, a random link may be chosen. Technically, the solution of this sub-problem provides a flow prioritization decision, as it defines to which loop $i$ the data rate of a link is assigned. The multipliers then play the role of prioritization factors. We can now define
\begin{equation}
	W_{mn}^i := \left[q_m^i - q_n^i\right]^+;\mathskip W_{mn} := \max_i \left\lbrace W_{mn}^i\right\rbrace.
\end{equation}

Using the optimal values and definitions, the stated problem has the simpler expression
\begin{equation}
	\max_{\mat{R}\in\mathset{C}}\sum\nolimits_{m\in\mathset{N}}\sum\nolimits_{n\in\mathset{N}}W_{mn}R_{mn}.
\end{equation}
This is a \gls{MAC} layer \gls{WSR} maximization, a problem that is well recognized in networking (e.g., \cite{Kobayashi2006,Christensen2008,Wong2008,Wong2009,Tan2011}). The explicit solution to this problem varies depending on the exact system properties, i.e., with the action set $\mathset{A}$ and link state set $\mathset{Q}$, respectively. However, it is well investigated under various system assumptions. We conclude that given $\vec{q}$, the value of $\Theta(\vec{q})$ can be obtained by solving a combination of optimal control, flow scheduling and \gls{WSR} maximization.

Given this development, we can choose the optimal multipliers by maximizing the dual function with a gradient ascent \cite{2004_Boyd,marino2016optimization}. As the optimal values of the Lagrange-variables satisfy $\lambda_i = q_{s_i}^i$ $\forall i$, it is only necessary to update $\vec{q}$. Using the formulation in \eqref{eq:final_lagrange}, this results in the update formula
\begin{equation}
	q_n^i[x+1]\! := \!\left[q_n^i[x] \!+\! \theta\!\left[\sum_{m\in\mathset{N}}\!\!R_{mn}^i  \!+\! \sum_{i\in\mathset{U}_n}\!\!r^i \!-\! \sum_{o\in\mathset{N}}\!\!R_{no}^i\right]\right]^+\label{eq:lagrangian_update}
\end{equation}
where $x\in\mathbb{N}_+$ is the iteration and $\theta > 0$ is a fixed step-size. 

From convex optimization theory \cite{2004_Boyd,marino2016optimization}, it is known that for appropriately chosen fixed step-sizes, gradient ascent algorithms converge towards a vicinity of the optimal solutions from arbitrary initial points. The size of the vicinity and convergence speed thereby depends on the step-size and smaller step-sizes result in a smaller vicinity but slower convergence. By comparing \eqref{eq:lagrangian_update} with \eqref{eq:mac_queue_evolution} and choosing $q_n^i[0] := \theta B_n^i[0]$ $\forall n,l$, it becomes clear that $q_n^i[x]:=\theta B_n^i[x]$, i.e., $q_n^i$ reflects the behavior of the respective \gls{MAC} layer queue. 

\vspace{-2mm}

\subsection{Optimal Networking}
For the optimal \gls{CC} and scheduling strategies $\psi$ and $\pi$, any value of $\mu_{n}^i$ satisfies the \gls{KKT} condition because the optimal multipliers are given by $\lambda_i \!=\! q_{s_i}^i$ $\forall i$. From the complementary slackness conditions, it must hold at optimality that 
\begin{equation}
	\lambda_i\left(r^i-\mu_{s_i}^i\right) = 0\mathskip \forall i.\label{eq:comp_slackness}
\end{equation}
We further deduce that at optimality, $\lambda_i \!>\! 0$ because $J^i(r^i)$ is decreasing in $r^i$. Let $\lambda_i \!=\! 0$ be an optimal multiplier, then $r^i$ could be ever increased to reduce the Lagrangian \eqref{eq:final_lagrange} further, which contradicts optimality of any $r^i$ and leads to infeasibility due to violation of the rate constraint. Hence $\lambda_i \!>\! 0$ must hold, leading to $\mu_{s_i}^i\!=\!r^i$ by \eqref{eq:comp_slackness}, i.e., the optimal admitted rate matches the injected one. Hence there is no need for CC in the given formulation. Indeed, as the sampling strategy is network-aware, it implicitly performs the task of \gls{CC}. Therefore, using an optimal sampling strategy, the best \gls{CC} policy is to pass through any traffic towards the \gls{MAC}-layer immediately.

\begin{algorithm}[t]
	\caption{Back-pressure Solution}
	\label{algo:backpressure_solution}
	\begin{algorithmic}[1] 
		\STATE Pre-design $\xi^i$ $\forall i$ using
		\STATE $\xi_k^{i,\ast}(Z_k^i)=-\vec{K}^i_\ast \;\mathbb{E}[\vec{x}_k^i|Z_k^i]$ according to \eqref{eq:opt_control_gain}-\eqref{eq:opt_estimator}\label{algo:control_law_design}
		\STATE Pre-design $M^i(\lambda)$ $\forall i$ (e.g., with value iteration \cite{2007Bertsekas})\label{algo:sampling_law_design}
		\STATE Choose $\theta>0$
		%		\STATE according to \cite{Molin2014}
		\FOR{$\tau = 1,...,\infty$}
		\FORALL{$i\in\mathset{L}$, $k \in \mathset{K}^i[\tau]$}
		\STATE $\vec{u}_k^i=\xi_k^{i,\ast}(Z_k^i)$ \label{algo:control_law}
		\STATE $\delta_k^i = \mathbbm{1}\lbrace  \|\vec{e}_{k|k-1}^i\|_2 > M^i(\theta B_{s_i}[\tau])\rbrace$\label{algo:sampling_law}
		\ENDFOR
		\FORALL{$(m,n)\in\mathset{N}\times\mathset{N}$}
		\FORALL{$i:(m,n)\in\mathset{Z}_i$}
		\STATE $W_{mn}^i[\tau] := \left[B_n^i[\tau]-B_{m}^i[\tau]\right]^+$\label{algo:diff_backlog}
		\ENDFOR
		\STATE $W_{mn}[\tau] := \max_i\lbrace W_{mn}^i[\tau]\rbrace$
		\STATE $i_{mn}^* := \arg\max_i\lbrace W_{mn}^i[\tau]\rbrace$
		\ENDFOR
		\STATE $\mat{A}[\tau] := \arg\max\limits_{\mat{A}\in\mathset{A}}\sum\limits_{m\in\mathset{N}}\sum\limits_{n\in\mathset{N}}W_{mn}R_{mn}(\mat{Q}[\tau],\mat{A})$\label{algo:bp_solution}
		\STATE Assign resulting rates to $i_{mn}^*$ $\forall (m,n)$
		\ENDFOR
	\end{algorithmic}
	
\end{algorithm}

\vspace{-2mm}

\subsection{Resulting Algorithm}
The optimization algorithm, i.e. Algorithm \ref{algo:backpressure_solution}, is a combination of certainty equivalent control, threshold based sampling and back-pressure scheduler 
\cite{2006_NOW_backpressure_book,2006_Lin_Schroff_Backpressure_Scheduling,4118456}: As shown in lines \ref{algo:control_law_design} and \ref{algo:control_law}, each loop uses the certainty equivalent control law in \eqref{eq:opt_control_gain}--\eqref{eq:opt_estimator}. Further, a threshold mapping $M^i:\mathbb{R}_+\mapsto \mathbb{R}_+$ from each Lagrange multiplier $\lambda_i$ to a threshold $M^i(\lambda_i)$ is designed. This can be done with value iteration \cite{2007Bertsekas}, using the error as system state and fixed~$\lambda_i$ as transmission cost, \cite{Molin2014}. In each control step, the samplers use a threshold policy (line \ref{algo:sampling_law}) where a scaled version of the \gls{MAC} back-log, $\theta B_{s_i}[\tau]$, is used as price value. On the network side each node $n\in\mathset{N}$ determines the  $W_{mn}^i[\tau]\!=\!\left[B_n^i[\tau]-B_{m}^i[\tau]\right]^+$ of all control loops whose traffic passes through them. The values are then used to solve a back-pressure scheduling problem in line \ref{algo:bp_solution} and the resulting rates are assigned to the loops with maximum differential back-log on each node. %Note that the step-size $\theta$ does not play a role in determining $W_{mn}^i[\tau]$ in line \ref{algo:diff_backlog} because all backlogs are equally scaled with the same value, such that the only effect of $\theta$ is scaling of the optimizing function in line \ref{algo:bp_solution}.%, hence they can be neglected. 

%\begin{remark}
%The above derivation considers optimization of feasible average rates but does not exactly tell how to realize these averages in a network. In general, the full $\mathset{C}$ is only achievable by averaging over different link actions in different time slots, so how to provide a desired expected rate is a non-trivial problem. However, while we derive the back-pressure algorithm from a dual optimization, it has also been derived and analyzed from Lyapunov optimization perspective \cite{2006_Lin_Schroff_Backpressure_Scheduling,2006_NOW_backpressure_book}. A key result is that if the weighted sum-rate problem is solved optimally in each time slot, then the network will be able to serve any admissible processes in $\mat{R}[\tau]$ and $\vec{r}[\tau] = [r^1[\tau],...,r^L[\tau]]^T$ for which $\vec{r} \in (\Lambda\setminus\partial\Lambda)$ holds \cite[Th. 4.5]{2006_NOW_backpressure_book}, i.e., any throughput in the interior of the transport capacity region. We can claim admissibility for $\mat{R}[\tau]$ due to our assumptions that $\mat{Q}[\tau]$ is a stationary process and that $\max_{\mathset{A},\mathset{Q}}\lVert\mat{R}(\mat{Q}[\tau],\mat{A}[\tau])\rVert_{\infty}\!<\!\infty$. Similarly, we can argue that each $r^i[\tau]$ is admissible as it is a random process with bounded conditional variance. Thus, our results can transfer to the case of randomly changing link states with stationary distribution while still resulting in a near-optimal solution.
%\end{remark}

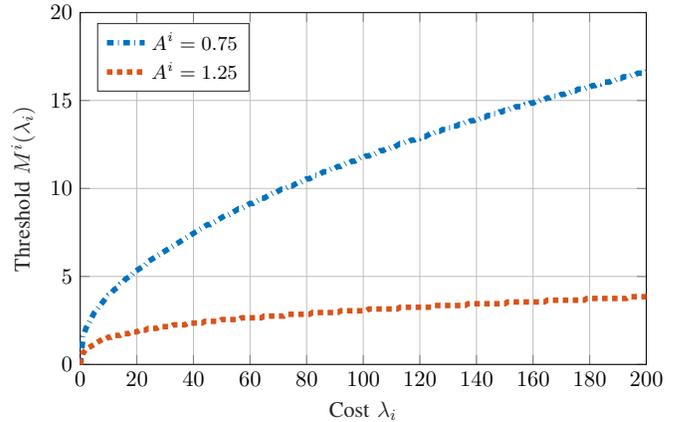
\begin{figure}[t]
	\centering\vspace{-3mm}
	\resizebox{\columnwidth}{!}{% This file was created by matlab2tikz.
\definecolor{mycolor1}{rgb}{0.00000,0.44700,0.74100}%
\definecolor{mycolor2}{rgb}{0.85000,0.32500,0.09800}%
\definecolor{mycolor3}{rgb}{0.92900,0.69400,0.12500}%
\definecolor{mycolor4}{rgb}{0.49400,0.18400,0.55600}%
\begin{tikzpicture}
\begin{axis}[%
width=9cm,
height=5.6cm,
scale only axis,
xmin=0,
xmax=200,
xlabel style={font=\color{white!15!black}},
xlabel={Cost $\lambda_i$},
ymin=0,
ymax=20,
ylabel style={font=\color{white!15!black}},
ylabel={Threshold $M^i(\lambda_i)$},
axis background/.style={fill=white},
xmajorgrids,
xminorgrids,
ymajorgrids,
legend style={at={(0.03,0.97)}, anchor=north west, legend cell align=left, align=left, draw=white!15!black,font=\small}
]
%\addplot [color=mycolor1, line width=3.0pt, mark size=3.5pt, mark=diamond, mark options={solid, mycolor1}]
\addplot [color=mycolor1, dashdotted, line width=2.5pt]
  table[row sep=crcr]{%
  	0	0\\
  	1	1.55\\
  	2	2.05\\
  	3	2.35\\
  	4	2.65\\
  	5	2.95\\
  	6	3.15\\
  	7	3.35\\
  	8	3.55\\
  	9	3.75\\
  	10	3.95\\
  	11	4.05\\
  	12	4.25\\
  	13	4.35\\
  	14	4.55\\
  	15	4.65\\
  	16	4.85\\
  	17	4.95\\
  	18	5.05\\
  	19	5.25\\
  	20	5.35\\
  	21	5.45\\
  	22	5.55\\
  	23	5.65\\
  	24	5.85\\
  	25	5.95\\
  	26	6.05\\
  	27	6.15\\
  	28	6.25\\
  	29	6.35\\
  	30	6.45\\
  	31	6.55\\
  	32	6.65\\
  	33	6.75\\
  	34	6.85\\
  	35	6.95\\
  	36	7.05\\
  	37	7.15\\
  	38	7.25\\
  	39	7.35\\
  	40	7.45\\
  	41	7.55\\
  	42	7.65\\
  	43	7.75\\
  	44	7.85\\
  	45	7.95\\
  	46	7.95\\
  	47	8.05\\
  	48	8.15\\
  	49	8.25\\
  	50	8.35\\
  	51	8.45\\
  	52	8.45\\
  	53	8.55\\
  	54	8.65\\
  	55	8.75\\
  	56	8.85\\
  	57	8.85\\
  	58	8.95\\
  	59	9.05\\
  	60	9.15\\
  	61	9.15\\
  	62	9.25\\
  	63	9.35\\
  	64	9.45\\
  	65	9.45\\
  	66	9.55\\
  	67	9.65\\
  	68	9.75\\
  	69	9.75\\
  	70	9.85\\
  	71	9.95\\
  	72	9.95\\
  	73	10.05\\
  	74	10.15\\
  	75	10.15\\
  	76	10.25\\
  	77	10.35\\
  	78	10.35\\
  	79	10.45\\
  	80	10.55\\
  	81	10.55\\
  	82	10.65\\
  	83	10.75\\
  	84	10.75\\
  	85	10.85\\
  	86	10.95\\
  	87	10.95\\
  	88	11.05\\
  	89	11.15\\
  	90	11.15\\
  	91	11.25\\
  	92	11.25\\
  	93	11.35\\
  	94	11.45\\
  	95	11.45\\
  	96	11.55\\
  	97	11.55\\
  	98	11.65\\
  	99	11.75\\
  	100	11.75\\
  	101	11.85\\
  	102	11.85\\
  	103	11.95\\
  	104	11.95\\
  	105	12.05\\
  	106	12.15\\
  	107	12.15\\
  	108	12.25\\
  	109	12.25\\
  	110	12.35\\
  	111	12.35\\
  	112	12.45\\
  	113	12.55\\
  	114	12.55\\
  	115	12.65\\
  	116	12.65\\
  	117	12.75\\
  	118	12.75\\
  	119	12.85\\
  	120	12.85\\
  	121	12.95\\
  	122	12.95\\
  	123	13.05\\
  	124	13.15\\
  	125	13.15\\
  	126	13.25\\
  	127	13.25\\
  	128	13.35\\
  	129	13.35\\
  	130	13.45\\
  	131	13.45\\
  	132	13.55\\
  	133	13.55\\
  	134	13.65\\
  	135	13.65\\
  	136	13.75\\
  	137	13.75\\
  	138	13.85\\
  	139	13.85\\
  	140	13.95\\
  	141	13.95\\
  	142	14.05\\
  	143	14.05\\
  	144	14.15\\
  	145	14.15\\
  	146	14.25\\
  	147	14.25\\
  	148	14.35\\
  	149	14.35\\
  	150	14.45\\
  	151	14.45\\
  	152	14.55\\
  	153	14.55\\
  	154	14.65\\
  	155	14.65\\
  	156	14.65\\
  	157	14.75\\
  	158	14.75\\
  	159	14.85\\
  	160	14.85\\
  	161	14.95\\
  	162	14.95\\
  	163	15.05\\
  	164	15.05\\
  	165	15.15\\
  	166	15.15\\
  	167	15.25\\
  	168	15.25\\
  	169	15.25\\
  	170	15.35\\
  	171	15.35\\
  	172	15.45\\
  	173	15.45\\
  	174	15.55\\
  	175	15.55\\
  	176	15.65\\
  	177	15.65\\
  	178	15.65\\
  	179	15.75\\
  	180	15.75\\
  	181	15.85\\
  	182	15.85\\
  	183	15.95\\
  	184	15.95\\
  	185	16.05\\
  	186	16.05\\
  	187	16.05\\
  	188	16.15\\
  	189	16.15\\
  	190	16.25\\
  	191	16.25\\
  	192	16.35\\
  	193	16.35\\
  	194	16.35\\
  	195	16.45\\
  	196	16.45\\
  	197	16.55\\
  	198	16.55\\
  	199	16.55\\
  	200	16.65\\
};
\addlegendentry{$A^i=0.75$}

%\addplot [color=mycolor2, dotted, line width=3.0pt, mark size=3.5pt, mark=triangle, mark options={solid, mycolor2}]
\addplot [color=mycolor2, dotted, line width=2.5pt]
  table[row sep=crcr]{%
  	0	0\\
  	1	0.65\\
  	2	0.85\\
  	3	0.949999999999999\\
  	4	1.05\\
  	5	1.15\\
  	6	1.25\\
  	7	1.35\\
  	8	1.45\\
  	9	1.45\\
  	10	1.55\\
  	11	1.55\\
  	12	1.65\\
  	13	1.65\\
  	14	1.65\\
  	15	1.75\\
  	16	1.75\\
  	17	1.75\\
  	18	1.85\\
  	19	1.85\\
  	20	1.85\\
  	21	1.95\\
  	22	1.95\\
  	23	1.95\\
  	24	2.05\\
  	25	2.05\\
  	26	2.05\\
  	27	2.05\\
  	28	2.15\\
  	29	2.15\\
  	30	2.15\\
  	31	2.15\\
  	32	2.15\\
  	33	2.25\\
  	34	2.25\\
  	35	2.25\\
  	36	2.25\\
  	37	2.25\\
  	38	2.35\\
  	39	2.35\\
  	40	2.35\\
  	41	2.35\\
  	42	2.35\\
  	43	2.35\\
  	44	2.45\\
  	45	2.45\\
  	46	2.45\\
  	47	2.45\\
  	48	2.45\\
  	49	2.45\\
  	50	2.55\\
  	51	2.55\\
  	52	2.55\\
  	53	2.55\\
  	54	2.55\\
  	55	2.55\\
  	56	2.55\\
  	57	2.65\\
  	58	2.65\\
  	59	2.65\\
  	60	2.65\\
  	61	2.65\\
  	62	2.65\\
  	63	2.65\\
  	64	2.65\\
  	65	2.75\\
  	66	2.75\\
  	67	2.75\\
  	68	2.75\\
  	69	2.75\\
  	70	2.75\\
  	71	2.75\\
  	72	2.75\\
  	73	2.85\\
  	74	2.85\\
  	75	2.85\\
  	76	2.85\\
  	77	2.85\\
  	78	2.85\\
  	79	2.85\\
  	80	2.85\\
  	81	2.85\\
  	82	2.95\\
  	83	2.95\\
  	84	2.95\\
  	85	2.95\\
  	86	2.95\\
  	87	2.95\\
  	88	2.95\\
  	89	2.95\\
  	90	2.95\\
  	91	2.95\\
  	92	3.05\\
  	93	3.05\\
  	94	3.05\\
  	95	3.05\\
  	96	3.05\\
  	97	3.05\\
  	98	3.05\\
  	99	3.05\\
  	100	3.05\\
  	101	3.05\\
  	102	3.15\\
  	103	3.15\\
  	104	3.15\\
  	105	3.15\\
  	106	3.15\\
  	107	3.15\\
  	108	3.15\\
  	109	3.15\\
  	110	3.15\\
  	111	3.15\\
  	112	3.15\\
  	113	3.25\\
  	114	3.25\\
  	115	3.25\\
  	116	3.25\\
  	117	3.25\\
  	118	3.25\\
  	119	3.25\\
  	120	3.25\\
  	121	3.25\\
  	122	3.25\\
  	123	3.25\\
  	124	3.25\\
  	125	3.35\\
  	126	3.35\\
  	127	3.35\\
  	128	3.35\\
  	129	3.35\\
  	130	3.35\\
  	131	3.35\\
  	132	3.35\\
  	133	3.35\\
  	134	3.35\\
  	135	3.35\\
  	136	3.35\\
  	137	3.45\\
  	138	3.45\\
  	139	3.45\\
  	140	3.45\\
  	141	3.45\\
  	142	3.45\\
  	143	3.45\\
  	144	3.45\\
  	145	3.45\\
  	146	3.45\\
  	147	3.45\\
  	148	3.45\\
  	149	3.45\\
  	150	3.45\\
  	151	3.55\\
  	152	3.55\\
  	153	3.55\\
  	154	3.55\\
  	155	3.55\\
  	156	3.55\\
  	157	3.55\\
  	158	3.55\\
  	159	3.55\\
  	160	3.55\\
  	161	3.55\\
  	162	3.55\\
  	163	3.55\\
  	164	3.55\\
  	165	3.65\\
  	166	3.65\\
  	167	3.65\\
  	168	3.65\\
  	169	3.65\\
  	170	3.65\\
  	171	3.65\\
  	172	3.65\\
  	173	3.65\\
  	174	3.65\\
  	175	3.65\\
  	176	3.65\\
  	177	3.65\\
  	178	3.65\\
  	179	3.75\\
  	180	3.75\\
  	181	3.75\\
  	182	3.75\\
  	183	3.75\\
  	184	3.75\\
  	185	3.75\\
  	186	3.75\\
  	187	3.75\\
  	188	3.75\\
  	189	3.75\\
  	190	3.75\\
  	191	3.75\\
  	192	3.75\\
  	193	3.75\\
  	194	3.85\\
  	195	3.85\\
  	196	3.85\\
  	197	3.85\\
  	198	3.85\\
  	199	3.85\\
  	200	3.85\\
};
\addlegendentry{$A^i=1.25$}

\end{axis}
\end{tikzpicture}%
}
	\vspace{-7mm}\caption{Sampling threshold $M^i$ as a function of $\lambda_i$, given $A^i \in \lbrace 0.75,1.25\rbrace$}\vspace{1mm}
	\label{fig:prices_vs_thresholds}
\end{figure}

\section{Simulation Results}
\label{sec:simulation}
We present a simulation study with the back-pressure solution from Algorithm \ref{algo:backpressure_solution}. We assume $|\mathset{L}|= \{2, \, 4, \, 6, \, \dots , \, 46\}$ control loops over a two-hop wireless network, in which all packets generated at the source nodes are forwarded to their destinations through a central base station node. Two transmission channels are available for each uplink and downlink hops to transport the data. We assume that the base station knows the queue backlog lengths and channel qualities through buffers status reports and channel sounding, based on which it enforces schedules according to a back-pressure scheduling law. 
To create diversity of CC mechanisms among loops, consider two classes of scalar plants with $A^i \!\in \!\{0.75, 1.25\}$. Let the number of stable plants ($A^i \!= \!0.75$) be equal with that of unstable ones ($A^i \!=\! 1.25$), $B^i\! =\! 1$, and $w^i \!\sim \!\mathset{N}(0,\,1)$, $\forall i \!\in \!\mathset{L}$. Packet size and channel quality is equal for all loops and each link can accommodate two users simultaneously. We consider equal sampling period among all control loops, and ten times faster communication $T_k = T^i  = 10 \cdot T_\tau,\, \forall i \in \mathset{L}$.

\begin{figure}[t]
	\resizebox{\columnwidth}{!}{% This file was created by matplotlib2tikz v0.7.4.
\begin{tikzpicture}

\definecolor{color0}{rgb}{0,0.447,0.741}
\definecolor{color1}{rgb}{0.85,0.325,0.098}

\begin{axis}[
width=9cm,
height=6cm,
legend cell align={left},
legend style={at={(0.03,0.03)}, anchor=south west, draw=white!80.0!black},
tick align=outside,
tick pos=left,
x grid style={white!69.01960784313725!black},
xlabel={\(\displaystyle L\)},
xmajorgrids,
xmin=-0.2, xmax=48.2,
xminorgrids,
xtick style={color=black},
y grid style={white!69.01960784313725!black},
ylabel={\(\displaystyle \overline{r}\)},
ymajorgrids,
ymin=-0.0494872190457963, ymax=1.04997558185932,
ytick style={color=black}
]
\path [draw=black, ultra thick]
(axis cs:2,1)
--(axis cs:2,1);

\path [draw=black, ultra thick]
(axis cs:4,1)
--(axis cs:4,1);

\path [draw=black, ultra thick]
(axis cs:6,1)
--(axis cs:6,1);

\path [draw=black, ultra thick]
(axis cs:8,1)
--(axis cs:8,1);

\path [draw=black, ultra thick]
(axis cs:10,1)
--(axis cs:10,1);

\path [draw=black, ultra thick]
(axis cs:12,1)
--(axis cs:12,1);

\path [draw=black, ultra thick]
(axis cs:14,1)
--(axis cs:14,1);

\path [draw=black, ultra thick]
(axis cs:16,1)
--(axis cs:16,1);

\path [draw=black, ultra thick]
(axis cs:18,1)
--(axis cs:18,1);

\path [draw=black, ultra thick]
(axis cs:20,1)
--(axis cs:20,1);

\path [draw=black, ultra thick]
(axis cs:22,0.909090219158507)
--(axis cs:22,0.909091370585083);

\path [draw=black, ultra thick]
(axis cs:24,0.833332691771844)
--(axis cs:24,0.833334256946105);

\path [draw=black, ultra thick]
(axis cs:26,0.769230108328966)
--(axis cs:26,0.76923137885052);

\path [draw=black, ultra thick]
(axis cs:28,0.714285004470476)
--(axis cs:28,0.714286610914139);

\path [draw=black, ultra thick]
(axis cs:30,0.666665747853214)
--(axis cs:30,0.666667175223709);

\path [draw=black, ultra thick]
(axis cs:32,0.624997892435017)
--(axis cs:32,0.624999620385496);

\path [draw=black, ultra thick]
(axis cs:34,0.588234562153869)
--(axis cs:34,0.588237283999977);

\path [draw=black, ultra thick]
(axis cs:36,0.555553048385981)
--(axis cs:36,0.555556515716583);

\path [draw=black, ultra thick]
(axis cs:38,0.526314526632139)
--(axis cs:38,0.526317806701194);

\path [draw=black, ultra thick]
(axis cs:40,0.499997619738876)
--(axis cs:40,0.500001559748304);

\path [draw=black, ultra thick]
(axis cs:42,0.476187987970346)
--(axis cs:42,0.476193781260423);

\path [draw=black, ultra thick]
(axis cs:44,0.454539956533058)
--(axis cs:44,0.454552146031045);

\path [draw=black, ultra thick]
(axis cs:46,0.500244181406764)
--(axis cs:46,0.500256598952211);

\path [draw=color0, ultra thick]
(axis cs:2,1)
--(axis cs:2,1);

\path [draw=color0, ultra thick]
(axis cs:4,1)
--(axis cs:4,1);

\path [draw=color0, ultra thick]
(axis cs:6,1)
--(axis cs:6,1);

\path [draw=color0, ultra thick]
(axis cs:8,1)
--(axis cs:8,1);

\path [draw=color0, ultra thick]
(axis cs:10,1)
--(axis cs:10,1);

\path [draw=color0, ultra thick]
(axis cs:12,1)
--(axis cs:12,1);

\path [draw=color0, ultra thick]
(axis cs:14,1)
--(axis cs:14,1);

\path [draw=color0, ultra thick]
(axis cs:16,1)
--(axis cs:16,1);

\path [draw=color0, ultra thick]
(axis cs:18,1)
--(axis cs:18,1);

\path [draw=color0, ultra thick]
(axis cs:20,1)
--(axis cs:20,1);

\path [draw=color0, ultra thick]
(axis cs:22,0.862831409548967)
--(axis cs:22,0.873769513527957);

\path [draw=color0, ultra thick]
(axis cs:24,0.751000665992489)
--(axis cs:24,0.760474821186999);

\path [draw=color0, ultra thick]
(axis cs:26,0.649486544337909)
--(axis cs:26,0.656667968482604);

\path [draw=color0, ultra thick]
(axis cs:28,0.565461754971035)
--(axis cs:28,0.572218296311017);

\path [draw=color0, ultra thick]
(axis cs:30,0.492859584698543)
--(axis cs:30,0.49704133837838);

\path [draw=color0, ultra thick]
(axis cs:32,0.428274835952897)
--(axis cs:32,0.431479779431718);

\path [draw=color0, ultra thick]
(axis cs:34,0.371148421851658)
--(axis cs:34,0.37323470635347);

\path [draw=color0, ultra thick]
(axis cs:36,0.321912891218779)
--(axis cs:36,0.322999724165837);

\path [draw=color0, ultra thick]
(axis cs:38,0.280857458184031)
--(axis cs:38,0.282333670021097);

\path [draw=color0, ultra thick]
(axis cs:40,0.240780413030177)
--(axis cs:40,0.241766971585208);

\path [draw=color0, ultra thick]
(axis cs:42,0.19323223599413)
--(axis cs:42,0.19387617426228);

\path [draw=color0, ultra thick]
(axis cs:44,0.130721379821231)
--(axis cs:44,0.131678363768512);

\path [draw=color0, ultra thick]
(axis cs:46,0.000488362813527371)
--(axis cs:46,0.000513197904421347);

\path [draw=color1, ultra thick]
(axis cs:2,1)
--(axis cs:2,1);

\path [draw=color1, ultra thick]
(axis cs:4,1)
--(axis cs:4,1);

\path [draw=color1, ultra thick]
(axis cs:6,1)
--(axis cs:6,1);

\path [draw=color1, ultra thick]
(axis cs:8,1)
--(axis cs:8,1);

\path [draw=color1, ultra thick]
(axis cs:10,1)
--(axis cs:10,1);

\path [draw=color1, ultra thick]
(axis cs:12,1)
--(axis cs:12,1);

\path [draw=color1, ultra thick]
(axis cs:14,1)
--(axis cs:14,1);

\path [draw=color1, ultra thick]
(axis cs:16,1)
--(axis cs:16,1);

\path [draw=color1, ultra thick]
(axis cs:18,1)
--(axis cs:18,1);

\path [draw=color1, ultra thick]
(axis cs:20,1)
--(axis cs:20,1);

\path [draw=color1, ultra thick]
(axis cs:22,0.944411917030859)
--(axis cs:22,0.955350339379397);

\path [draw=color1, ultra thick]
(axis cs:24,0.906191788393354)
--(axis cs:24,0.915666621863056);

\path [draw=color1, ultra thick]
(axis cs:26,0.881793487720466)
--(axis cs:26,0.888974973817995);

\path [draw=color1, ultra thick]
(axis cs:28,0.856353010492523)
--(axis cs:28,0.863110168994657);

\path [draw=color1, ultra thick]
(axis cs:30,0.836291706098)
--(axis cs:30,0.840473216978923);

\path [draw=color1, ultra thick]
(axis cs:32,0.818517789609539)
--(axis cs:32,0.821722620646871);

\path [draw=color1, ultra thick]
(axis cs:34,0.803238128864632)
--(axis cs:34,0.805322435237932);

\path [draw=color1, ultra thick]
(axis cs:36,0.788109695341651)
--(axis cs:36,0.789196817478862);

\path [draw=color1, ultra thick]
(axis cs:38,0.770298933334128)
--(axis cs:38,0.77177460512741);

\path [draw=color1, ultra thick]
(axis cs:40,0.758231500956293)
--(axis cs:40,0.759219473402682);

\path [draw=color1, ultra thick]
(axis cs:42,0.758505189599596)
--(axis cs:42,0.759149938605532);

\path [draw=color1, ultra thick]
(axis cs:44,0.777411096852904)
--(axis cs:44,0.778373364685558);

\path [draw=color1, ultra thick]
(axis cs:46,1)
--(axis cs:46,1);

\addplot [ultra thick, black]
table {%
2 1
4 1
6 1
8 1
10 1
12 1
14 1
16 1
18 1
20 1
22 0.909090794871795
24 0.833333474358974
26 0.769230743589743
28 0.714285807692308
30 0.666666461538462
32 0.624998756410256
34 0.588235923076923
36 0.555554782051282
38 0.526316166666667
40 0.49999958974359
42 0.476190884615385
44 0.454546051282051
46 0.500250390179487
};
\addlegendentry{$\overline{r}$}
\addplot [ultra thick, color0, dashed]
table {%
2 1
4 1
6 1
8 1
10 1
12 1
14 1
16 1
18 1
20 1
22 0.868300461538462
24 0.755737743589744
26 0.653077256410256
28 0.568840025641026
30 0.494950461538462
32 0.429877307692308
34 0.372191564102564
36 0.322456307692308
38 0.281595564102564
40 0.241273692307692
42 0.193554205128205
44 0.131199871794872
46 0.000500780358974359
};
\addlegendentry{$\overline{r}_{0.75}$}
\addplot [ultra thick, color1, dotted]
table {%
2 1
4 1
6 1
8 1
10 1
12 1
14 1
16 1
18 1
20 1
22 0.949881128205128
24 0.910929205128205
26 0.885384230769231
28 0.85973158974359
30 0.838382461538462
32 0.820120205128205
34 0.804280282051282
36 0.788653256410256
38 0.771036769230769
40 0.758725487179487
42 0.758827564102564
44 0.777892230769231
46 1
};
\addlegendentry{$\overline{r}_{1.25}$}
\end{axis}

\end{tikzpicture}}
	\vspace{-6mm}\caption{Solid line: average packet injection rate per system, i.e., $\overline{r}$. Dashed lines $\overline{r}_{0.75}$ and $\overline{r}_{1.25}$: average rate of stable and unstable plants, respectively. %Vertical error bars represent $95\%$ confidence intervals.
	}\vspace{-1mm}
	\label{fig:rate}
\end{figure}
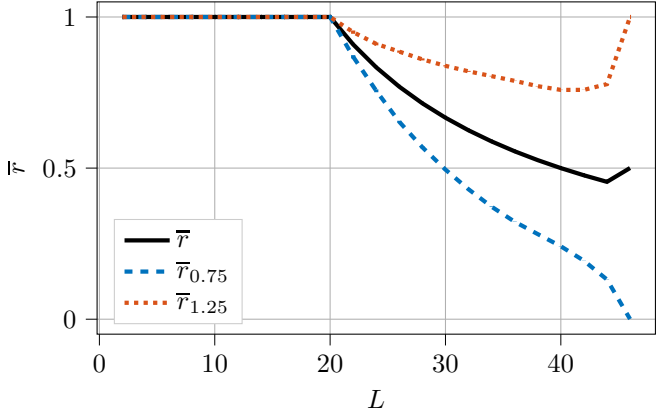

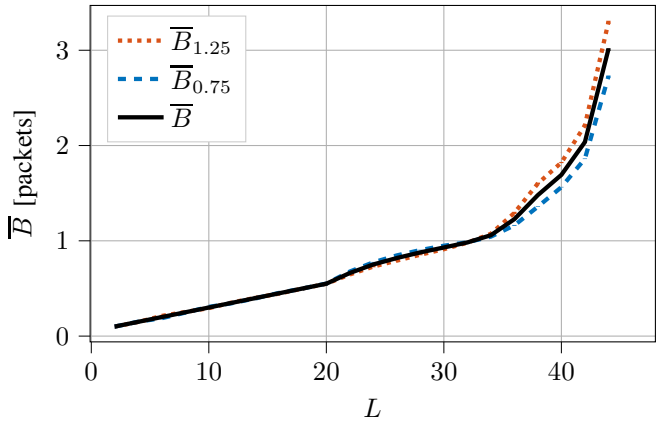
\begin{figure}[t]
	\resizebox{\columnwidth}{!}{% This file was created by matplotlib2tikz v0.7.4.
\begin{tikzpicture}

\definecolor{color0}{rgb}{0.85,0.325,0.098}
\definecolor{color1}{rgb}{0,0.447,0.741}

\begin{axis}[
width=9cm,
height=6cm,
legend cell align={left},
legend style={at={(0.03,0.97)}, anchor=north west, draw=white!80.0!black},
tick align=outside,
tick pos=left,
x grid style={white!69.01960784313725!black},
xlabel={\(\displaystyle L\)},
xmajorgrids,
xmin=-0.1, xmax=48,
xminorgrids,
xtick style={color=black},
y grid style={white!69.01960784313725!black},
ylabel={\(\displaystyle \overline{B}\) [packets]},
ymajorgrids,
ymin=-0.0604815077422007, ymax=3.47011166258621,
ytick style={color=black}
]
\path [draw=color0, ultra thick]
(axis cs:2,0.1)
--(axis cs:2,0.1);

\path [draw=color0, ultra thick]
(axis cs:4,0.13900082164184)
--(axis cs:4,0.157914973229955);

\path [draw=color0, ultra thick]
(axis cs:6,0.201356247716694)
--(axis cs:6,0.222832726642281);

\path [draw=color0, ultra thick]
(axis cs:8,0.242503404576903)
--(axis cs:8,0.264393210807712);

\path [draw=color0, ultra thick]
(axis cs:10,0.281223805381973)
--(axis cs:10,0.305095220259053);

\path [draw=color0, ultra thick]
(axis cs:12,0.340686927328919)
--(axis cs:12,0.367580098312106);

\path [draw=color0, ultra thick]
(axis cs:14,0.382597180257399)
--(axis cs:14,0.422543127434909);

\path [draw=color0, ultra thick]
(axis cs:16,0.431361666653047)
--(axis cs:16,0.461020794885414);

\path [draw=color0, ultra thick]
(axis cs:18,0.483303909045745)
--(axis cs:18,0.516311680697845);

\path [draw=color0, ultra thick]
(axis cs:20,0.536181319088848)
--(axis cs:20,0.561453962962434);

\path [draw=color0, ultra thick]
(axis cs:22,0.636562719106618)
--(axis cs:22,0.657196255252356);

\path [draw=color0, ultra thick]
(axis cs:24,0.721799487897853)
--(axis cs:24,0.737802922358558);

\path [draw=color0, ultra thick]
(axis cs:26,0.78636387401031)
--(axis cs:26,0.797930741374305);

\path [draw=color0, ultra thick]
(axis cs:28,0.848268034017809)
--(axis cs:28,0.860278478802703);

\path [draw=color0, ultra thick]
(axis cs:30,0.905822816673478)
--(axis cs:30,0.917736619223958);

\path [draw=color0, ultra thick]
(axis cs:32,0.974063510487433)
--(axis cs:32,0.986657617717695);

\path [draw=color0, ultra thick]
(axis cs:34,1.06633259796845)
--(axis cs:34,1.0813412481854);

\path [draw=color0, ultra thick]
(axis cs:36,1.28121882910522)
--(axis cs:36,1.30444373499735);

\path [draw=color0, ultra thick]
(axis cs:38,1.59153579532319)
--(axis cs:38,1.61105240980501);

\path [draw=color0, ultra thick]
(axis cs:40,1.81521868546648)
--(axis cs:40,1.82785157094378);

\path [draw=color0, ultra thick]
(axis cs:42,2.20213150842093)
--(axis cs:42,2.22007618388677);

\path [draw=color0, ultra thick]
(axis cs:44,3.2775344605406)
--(axis cs:44,3.30963015484401);

\path [draw=color1, ultra thick]
(axis cs:2,0.1)
--(axis cs:2,0.1);

\path [draw=color1, ultra thick]
(axis cs:4,0.142085026770045)
--(axis cs:4,0.16099917835816);

\path [draw=color1, ultra thick]
(axis cs:6,0.177167265703621)
--(axis cs:6,0.198643708655353);

\path [draw=color1, ultra thick]
(axis cs:8,0.235606845703929)
--(axis cs:8,0.257496641475558);

\path [draw=color1, ultra thick]
(axis cs:10,0.294904702944229)
--(axis cs:10,0.31877606628654);

\path [draw=color1, ultra thick]
(axis cs:12,0.332419940676471)
--(axis cs:12,0.359313084964555);

\path [draw=color1, ultra thick]
(axis cs:14,0.377456848713218)
--(axis cs:14,0.417402741030372);

\path [draw=color1, ultra thick]
(axis cs:16,0.438979217359273)
--(axis cs:16,0.468638321102266);

\path [draw=color1, ultra thick]
(axis cs:18,0.483688319302156)
--(axis cs:18,0.516696090954255);

\path [draw=color1, ultra thick]
(axis cs:20,0.538546010074339)
--(axis cs:20,0.563818605310276);

\path [draw=color1, ultra thick]
(axis cs:22,0.670795447260513)
--(axis cs:22,0.684295373252308);

\path [draw=color1, ultra thick]
(axis cs:24,0.770258041484927)
--(axis cs:24,0.778480625181739);

\path [draw=color1, ultra thick]
(axis cs:26,0.843214059315308)
--(axis cs:26,0.850631017607769);

\path [draw=color1, ultra thick]
(axis cs:28,0.899629423829991)
--(axis cs:28,0.906214319759753);

\path [draw=color1, ultra thick]
(axis cs:30,0.94416018928868)
--(axis cs:30,0.952650528660038);

\path [draw=color1, ultra thick]
(axis cs:32,0.980103150525093)
--(axis cs:32,0.989835823833881);

\path [draw=color1, ultra thick]
(axis cs:34,1.03894769012233)
--(axis cs:34,1.04898513039049);

\path [draw=color1, ultra thick]
(axis cs:36,1.15763426059202)
--(axis cs:36,1.16639804710029);

\path [draw=color1, ultra thick]
(axis cs:38,1.3552669417973)
--(axis cs:38,1.36184741717705);

\path [draw=color1, ultra thick]
(axis cs:40,1.5603338261373)
--(axis cs:40,1.56597540463193);

\path [draw=color1, ultra thick]
(axis cs:42,1.85797468602158)
--(axis cs:42,1.86400890372201);

\path [draw=color1, ultra thick]
(axis cs:44,2.71265964840333)
--(axis cs:44,2.73484445416077);

\path [draw=black, ultra thick]
(axis cs:2,0.1)
--(axis cs:2,0.1);

\path [draw=black, ultra thick]
(axis cs:4,0.15)
--(axis cs:4,0.15);

\path [draw=black, ultra thick]
(axis cs:6,0.199999961225716)
--(axis cs:6,0.200000013133259);

\path [draw=black, ultra thick]
(axis cs:8,0.24999998942302)
--(axis cs:8,0.250000061859032);

\path [draw=black, ultra thick]
(axis cs:10,0.299999876281937)
--(axis cs:10,0.300000021153961);

\path [draw=black, ultra thick]
(axis cs:12,0.349999986866741)
--(axis cs:12,0.350000038774284);

\path [draw=black, ultra thick]
(axis cs:14,0.399999938140969)
--(axis cs:14,0.40000001057698);

\path [draw=black, ultra thick]
(axis cs:16,0.449999962816011)
--(axis cs:16,0.450000037183989);

\path [draw=black, ultra thick]
(axis cs:18,0.5)
--(axis cs:18,0.5);

\path [draw=black, ultra thick]
(axis cs:20,0.549999910507255)
--(axis cs:20,0.550000038210694);

\path [draw=black, ultra thick]
(axis cs:22,0.657096602668356)
--(axis cs:22,0.667328294767542);

\path [draw=black, ultra thick]
(axis cs:24,0.747879529869167)
--(axis cs:24,0.756291008592372);

\path [draw=black, ultra thick]
(axis cs:26,0.816515035977739)
--(axis cs:26,0.822554810176108);

\path [draw=black, ultra thick]
(axis cs:28,0.875920566342502)
--(axis cs:28,0.881274561862626);

\path [draw=black, ultra thick]
(axis cs:30,0.928517945400392)
--(axis cs:30,0.931667131522685);

\path [draw=black, ultra thick]
(axis cs:32,0.981219697927423)
--(axis cs:32,0.984110353354628);

\path [draw=black, ultra thick]
(axis cs:34,1.05726597501598)
--(axis cs:34,1.06053735831736);

\path [draw=black, ultra thick]
(axis cs:36,1.22354452192604)
--(axis cs:36,1.2313029139714);

\path [draw=black, ultra thick]
(axis cs:38,1.476233627051)
--(axis cs:38,1.48361765500028);

\path [draw=black, ultra thick]
(axis cs:40,1.6900998814963)
--(axis cs:40,1.69458986209344);

\path [draw=black, ultra thick]
(axis cs:42,2.03131663333331)
--(axis cs:42,2.04077900769233);

\path [draw=black, ultra thick]
(axis cs:44,2.9953063281736)
--(axis cs:44,3.02202803080076);

\addplot [ultra thick, color0, dotted]
table {%
2 0.1
4 0.148457897435897
6 0.212094487179487
8 0.253448307692308
10 0.293159512820513
12 0.354133512820513
14 0.402570153846154
16 0.446191230769231
18 0.499807794871795
20 0.548817641025641
22 0.646879487179487
24 0.729801205128205
26 0.792147307692308
28 0.854273256410256
30 0.911779717948718
32 0.980360564102564
34 1.07383692307692
36 1.29283128205128
38 1.6012941025641
40 1.82153512820513
42 2.21110384615385
44 3.29358230769231
};
\addlegendentry{$\overline{B}_{1.25}$}
\addplot [ultra thick, color1, dashed]
table {%
2 0.1
4 0.151542102564103
6 0.187905487179487
8 0.246551743589744
10 0.306840384615385
12 0.345866512820513
14 0.397429794871795
16 0.453808769230769
18 0.500192205128205
20 0.551182307692308
22 0.67754541025641
24 0.774369333333333
26 0.846922538461539
28 0.902921871794872
30 0.948405358974359
32 0.984969487179487
34 1.04396641025641
36 1.16201615384615
38 1.35855717948718
40 1.56315461538462
42 1.8609917948718
44 2.72375205128205
};
\addlegendentry{$\overline{B}_{0.75}$}
\addplot [ultra thick, black]
table {%
2 0.1
4 0.15
6 0.199999987179487
8 0.250000025641026
10 0.299999948717949
12 0.350000012820513
14 0.399999974358974
16 0.45
18 0.5
20 0.549999974358975
22 0.662212448717949
24 0.752085269230769
26 0.819534923076923
28 0.878597564102564
30 0.930092538461538
32 0.982665025641026
34 1.05890166666667
36 1.22742371794872
38 1.47992564102564
40 1.69234487179487
42 2.03604782051282
44 3.00866717948718
};
\addlegendentry{$\overline{B}$}
\end{axis}

\end{tikzpicture}}
	\vspace{-6mm}\caption{Solid line: average queue backlog of the MAC buffer per system, i.e., $\overline{B}$. Dashed lines $\overline{B}_{0.75}$ and $\overline{B}_{1.25}$: average queue backlog of stable and unstable plants, respectively. Vertical error bars with $95\%$ confidence intervals.}\vspace{-1mm}
	\label{fig:queueBacklog}
\end{figure}

Fig. \ref{fig:prices_vs_thresholds} illustrates the threshold values for both classes of loops for $Q_x^i\! =\! 1.0$ and $Q_u^i \!=\! 0, \forall i$. From (\ref{eq:opt_control_gain}) together with $Q_u^i \!=\! 0$, the optimal control becomes a deadbeat controller with the optimal gain $K^i_{\ast}=A^i$. The values are obtained by using the value iteration to solve an average cost-per-stage problem for each fixed $\lambda_i$, following \cite{Molin2014}. Note that as shown in Section~\ref{subsec:DualDecomposition}, threshold values of the $i$-th loop $\lambda_i$ are proportional to the current backlog in the MAC layer queue, i.e., $\lambda_i = q_{s_i}^i = \theta B_{s_i}^i$. A higher queue backlog leads to a higher threshold value which reduces the chance of a packet injection at the source, effectively leading to a CC mechanism. Further, it is observed that rate reduction is more aggressive for stable plants due to the increased threshold.

\begin{figure}[t]
	\resizebox{\columnwidth}{!}{% This file was created by matplotlib2tikz v0.7.4.
\begin{tikzpicture}

\definecolor{color0}{rgb}{0.85,0.325,0.098}
\definecolor{color1}{rgb}{0,0.447,0.741}

\begin{axis}[
width=9cm,
height=6cm,
legend cell align={left},
legend style={at={(0.03,0.97)}, anchor=north west, draw=white!80.0!black},
tick align=outside,
tick pos=left,
x grid style={white!69.01960784313725!black},
xlabel={\(\displaystyle L\)},
xmajorgrids,
xmin=-0.1, xmax=48,
xminorgrids,
xtick style={color=black},
y grid style={white!69.01960784313725!black},
ylabel={Average Delay [time-steps]},
ymajorgrids,
ymin=-1.02318237995227, ymax=21.4868299789977,
yminorgrids,
ytick style={color=black}
]
\path [draw=color0, ultra thick]
(axis cs:2,0)
--(axis cs:2,0);

\path [draw=color0, ultra thick]
(axis cs:4,0)
--(axis cs:4,0);

\path [draw=color0, ultra thick]
(axis cs:6,0)
--(axis cs:6,0);

\path [draw=color0, ultra thick]
(axis cs:8,0)
--(axis cs:8,0);

\path [draw=color0, ultra thick]
(axis cs:10,0)
--(axis cs:10,0);

\path [draw=color0, ultra thick]
(axis cs:12,0)
--(axis cs:12,0);

\path [draw=color0, ultra thick]
(axis cs:14,0)
--(axis cs:14,0);

\path [draw=color0, ultra thick]
(axis cs:16,0)
--(axis cs:16,0);

\path [draw=color0, ultra thick]
(axis cs:18,0)
--(axis cs:18,0);

\path [draw=color0, ultra thick]
(axis cs:20,0.0968335533999327)
--(axis cs:20,0.129182964548785);

\path [draw=color0, ultra thick]
(axis cs:22,0.238136006076429)
--(axis cs:22,0.279051378538956);

\path [draw=color0, ultra thick]
(axis cs:24,0.368444294655346)
--(axis cs:24,0.406368372011321);

\path [draw=color0, ultra thick]
(axis cs:26,0.467088445965466)
--(axis cs:26,0.492360938649918);

\path [draw=color0, ultra thick]
(axis cs:28,0.558899102227928)
--(axis cs:28,0.58285161572079);

\path [draw=color0, ultra thick]
(axis cs:30,0.647198618982646)
--(axis cs:30,0.662183688709661);

\path [draw=color0, ultra thick]
(axis cs:32,0.74391798663076)
--(axis cs:32,0.759183141574369);

\path [draw=color0, ultra thick]
(axis cs:34,0.881850128681335)
--(axis cs:34,0.899123461062255);

\path [draw=color0, ultra thick]
(axis cs:36,1.18621859848659)
--(axis cs:36,1.21870345279546);

\path [draw=color0, ultra thick]
(axis cs:38,1.62713551266099)
--(axis cs:38,1.6563639745185);

\path [draw=color0, ultra thick]
(axis cs:40,1.94980536260512)
--(axis cs:40,1.96784899636924);

\path [draw=color0, ultra thick]
(axis cs:42,2.45999815449492)
--(axis cs:42,2.48180851217174);

\path [draw=color0, ultra thick]
(axis cs:44,3.7710177233446)
--(axis cs:44,3.80742894332207);

\path [draw=color1, ultra thick]
(axis cs:2,0)
--(axis cs:2,0);

\path [draw=color1, ultra thick]
(axis cs:4,0)
--(axis cs:4,0);

\path [draw=color1, ultra thick]
(axis cs:6,0)
--(axis cs:6,0);

\path [draw=color1, ultra thick]
(axis cs:8,0)
--(axis cs:8,0);

\path [draw=color1, ultra thick]
(axis cs:10,0)
--(axis cs:10,0);

\path [draw=color1, ultra thick]
(axis cs:12,0)
--(axis cs:12,0);

\path [draw=color1, ultra thick]
(axis cs:14,0)
--(axis cs:14,0);

\path [draw=color1, ultra thick]
(axis cs:16,0)
--(axis cs:16,0);

\path [draw=color1, ultra thick]
(axis cs:18,0)
--(axis cs:18,0);

\path [draw=color1, ultra thick]
(axis cs:20,0.0708170504239844)
--(axis cs:20,0.103166423422169);

\path [draw=color1, ultra thick]
(axis cs:22,0.3136603007379)
--(axis cs:22,0.341202417210818);

\path [draw=color1, ultra thick]
(axis cs:24,0.545436047783648)
--(axis cs:24,0.577311029139429);

\path [draw=color1, ultra thick]
(axis cs:26,0.82839672944422)
--(axis cs:26,0.848464552607062);

\path [draw=color1, ultra thick]
(axis cs:28,1.11741075209148)
--(axis cs:28,1.14705181201109);

\path [draw=color1, ultra thick]
(axis cs:30,1.45685647420079)
--(axis cs:30,1.48063019246587);

\path [draw=color1, ultra thick]
(axis cs:32,1.83658884077974)
--(axis cs:32,1.85983577460488);

\path [draw=color1, ultra thick]
(axis cs:34,2.34006098858074)
--(axis cs:34,2.36488413962439);

\path [draw=color1, ultra thick]
(axis cs:36,3.1093487412)
--(axis cs:36,3.15103895110769);

\path [draw=color1, ultra thick]
(axis cs:38,4.32648949520058)
--(axis cs:38,4.37637255608147);

\path [draw=color1, ultra thick]
(axis cs:40,5.99741722993582)
--(axis cs:40,6.04084020596162);

\path [draw=color1, ultra thick]
(axis cs:42,9.11956345314399)
--(axis cs:42,9.17326064942011);

\path [draw=color1, ultra thick]
(axis cs:44,20.1486754778777)
--(axis cs:44,20.4636475990454);

\path [draw=black, ultra thick]
(axis cs:2,0)
--(axis cs:2,0);

\path [draw=black, ultra thick]
(axis cs:4,0)
--(axis cs:4,0);

\path [draw=black, ultra thick]
(axis cs:6,0)
--(axis cs:6,0);

\path [draw=black, ultra thick]
(axis cs:8,0)
--(axis cs:8,0);

\path [draw=black, ultra thick]
(axis cs:10,0)
--(axis cs:10,0);

\path [draw=black, ultra thick]
(axis cs:12,0)
--(axis cs:12,0);

\path [draw=black, ultra thick]
(axis cs:14,0)
--(axis cs:14,0);

\path [draw=black, ultra thick]
(axis cs:16,0)
--(axis cs:16,0);

\path [draw=black, ultra thick]
(axis cs:18,0)
--(axis cs:18,0);

\path [draw=black, ultra thick]
(axis cs:20,0.0999998233483222)
--(axis cs:20,0.100000172549114);

\path [draw=black, ultra thick]
(axis cs:22,0.288247829565753)
--(axis cs:22,0.297777221716299);

\path [draw=black, ultra thick]
(axis cs:24,0.470955310065911)
--(axis cs:24,0.47782456172896);

\path [draw=black, ultra thick]
(axis cs:26,0.655608098993012)
--(axis cs:26,0.662547234340321);

\path [draw=black, ultra thick]
(axis cs:28,0.847896006018913)
--(axis cs:28,0.855210635006728);

\path [draw=black, ultra thick]
(axis cs:30,1.05861535584438)
--(axis cs:30,1.06481913133511);

\path [draw=black, ultra thick]
(axis cs:32,1.29756503589588)
--(axis cs:32,1.30219783589899);

\path [draw=black, ultra thick]
(axis cs:34,1.61925250407232)
--(axis cs:34,1.62370685490204);

\path [draw=black, ultra thick]
(axis cs:36,2.16275189512032)
--(axis cs:36,2.16990297667455);

\path [draw=black, ultra thick]
(axis cs:38,2.98968812454686)
--(axis cs:38,3.0034926446839);

\path [draw=black, ultra thick]
(axis cs:40,3.98168827148152)
--(axis cs:40,3.99626762595438);

\path [draw=black, ultra thick]
(axis cs:42,5.7928876868986)
--(axis cs:42,5.82442769771679);

\path [draw=black, ultra thick]
(axis cs:44,11.960545527552)
--(axis cs:44,12.1348393442428);

\addplot [ultra thick, color0, dotted]
table {%
2 0
4 0
6 0
8 0
10 0
12 0
14 0
16 0
18 0
20 0.113008258974359
22 0.258593692307692
24 0.387406333333333
26 0.479724692307692
28 0.570875358974359
30 0.654691153846154
32 0.751550564102564
34 0.890486794871795
36 1.20246102564103
38 1.64174974358974
40 1.95882717948718
42 2.47090333333333
44 3.78922333333333
};
\addlegendentry{$\overline{d}_{1.25}$}
\addplot [ultra thick, color1, dashed]
table {%
2 0
4 0
6 0
8 0
10 0
12 0
14 0
16 0
18 0
20 0.0869917369230769
22 0.327431358974359
24 0.561373538461538
26 0.838430641025641
28 1.13223128205128
30 1.46874333333333
32 1.84821230769231
34 2.35247256410256
36 3.13019384615385
38 4.35143102564103
40 6.01912871794872
42 9.14641205128205
44 20.3061615384615
};
\addlegendentry{$\overline{d}_{0.75}$}
\addplot [ultra thick, black]
table {%
2 0
4 0
6 0
8 0
10 0
12 0
14 0
16 0
18 0
20 0.0999999979487179
22 0.293012525641026
24 0.474389935897436
26 0.659077666666667
28 0.85155332051282
30 1.06171724358974
32 1.29988143589744
34 1.62147967948718
36 2.16632743589744
38 2.99659038461538
40 3.98897794871795
42 5.80865769230769
44 12.0476924358974
};
\addlegendentry{$\overline{d}$}
\end{axis}

\end{tikzpicture}}
	\vspace{-6mm}\caption{The solid line illustrates the average delay per system, i.e., $\overline{d}$. Dashed lines $\overline{d}_{0.75}$ and $\overline{d}_{1.25}$ show the delay of stable and unstable plants, respectively. Vertical error bars represent $95\%$ confidence intervals.}
	\label{fig:delay}
\end{figure}
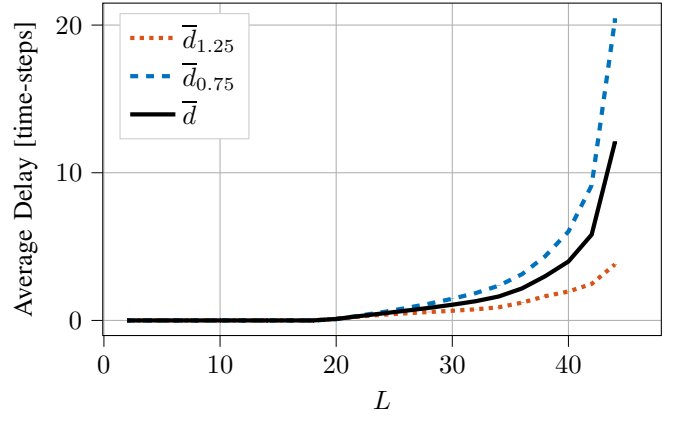

\begin{figure}[t]
	\resizebox{\columnwidth}{!}{% This file was created by matplotlib2tikz v0.7.4.
\begin{tikzpicture}

\definecolor{color0}{rgb}{0.85,0.325,0.098}
\definecolor{color1}{rgb}{0,0.447,0.741}

\begin{axis}[
width=9cm,
height=6cm,
legend cell align={left},
legend style={at={(0.03,0.97)}, anchor=north west, draw=white!80.0!black},
tick align=outside,
tick pos=left,
x grid style={white!69.01960784313725!black},
xlabel={\(\displaystyle L\)},
xmajorgrids,
xmin=-0.1, xmax=48,
xminorgrids,
xtick style={color=black},
y grid style={white!69.01960784313725!black},
ylabel={Average Cost \(\displaystyle \overline{J}\)},
ymajorgrids,
ymin=0.195304945241105, ymax=17.7585255554211,
ytick style={color=black}
]
\path [draw=color0, ultra thick]
(axis cs:2,0.997073185306534)
--(axis cs:2,1.00284378905244);

\path [draw=color0, ultra thick]
(axis cs:4,1.00163545768633)
--(axis cs:4,1.00497659359572);

\path [draw=color0, ultra thick]
(axis cs:6,1.00263694801488)
--(axis cs:6,1.0065681289082);

\path [draw=color0, ultra thick]
(axis cs:8,0.997189496361461)
--(axis cs:8,0.999630195946232);

\path [draw=color0, ultra thick]
(axis cs:10,0.993633154794742)
--(axis cs:10,0.99676171700013);

\path [draw=color0, ultra thick]
(axis cs:12,0.995901663717526)
--(axis cs:12,0.998910541410679);

\path [draw=color0, ultra thick]
(axis cs:14,0.996275464944699)
--(axis cs:14,0.998452176080942);

\path [draw=color0, ultra thick]
(axis cs:16,0.997729964251327)
--(axis cs:16,0.99969952292816);

\path [draw=color0, ultra thick]
(axis cs:18,0.996952208669768)
--(axis cs:18,0.999305227227668);

\path [draw=color0, ultra thick]
(axis cs:20,1.15496222744811)
--(axis cs:20,1.20559674691087);

\path [draw=color0, ultra thick]
(axis cs:22,1.37563934973668)
--(axis cs:22,1.4359319323146);

\path [draw=color0, ultra thick]
(axis cs:24,1.55719560804187)
--(axis cs:24,1.60581567400941);

\path [draw=color0, ultra thick]
(axis cs:26,1.6743657256707)
--(axis cs:26,1.70029786407289);

\path [draw=color0, ultra thick]
(axis cs:28,1.75402357447016)
--(axis cs:28,1.77242155373497);

\path [draw=color0, ultra thick]
(axis cs:30,1.82391934201413)
--(axis cs:30,1.83674629901151);

\path [draw=color0, ultra thick]
(axis cs:32,1.91353636003779)
--(axis cs:32,1.93359184509042);

\path [draw=color0, ultra thick]
(axis cs:34,2.12020475778552)
--(axis cs:34,2.16102498580423);

\path [draw=color0, ultra thick]
(axis cs:36,2.82209862285995)
--(axis cs:36,2.90936855662722);

\path [draw=color0, ultra thick]
(axis cs:38,4.00621689504805)
--(axis cs:38,4.09227951520836);

\path [draw=color0, ultra thick]
(axis cs:40,4.98499616502226)
--(axis cs:40,5.05151152728543);

\path [draw=color0, ultra thick]
(axis cs:42,6.99933557464389)
--(axis cs:42,7.09914339971509);

\path [draw=color0, ultra thick]
(axis cs:44,15.9706744490043)
--(axis cs:44,16.9601973458675);

\path [draw=color1, ultra thick]
(axis cs:2,0.996606689339775)
--(axis cs:2,1.00069982348074);

\path [draw=color1, ultra thick]
(axis cs:4,0.99746982890916)
--(axis cs:4,1.00143232493699);

\path [draw=color1, ultra thick]
(axis cs:6,0.998317384950097)
--(axis cs:6,1.00149482017811);

\path [draw=color1, ultra thick]
(axis cs:8,0.996890724495522)
--(axis cs:8,0.99971932678653);

\path [draw=color1, ultra thick]
(axis cs:10,0.997961398644278)
--(axis cs:10,1.00043552443264);

\path [draw=color1, ultra thick]
(axis cs:12,0.998484184316154)
--(axis cs:12,1.00034709773513);

\path [draw=color1, ultra thick]
(axis cs:14,0.996689298728594)
--(axis cs:14,0.998745880758586);

\path [draw=color1, ultra thick]
(axis cs:16,0.997619358464136)
--(axis cs:16,0.999377769740992);

\path [draw=color1, ultra thick]
(axis cs:18,0.996117031804914)
--(axis cs:18,0.998517327169445);

\path [draw=color1, ultra thick]
(axis cs:20,1.04091149893503)
--(axis cs:20,1.05887496260343);

\path [draw=color1, ultra thick]
(axis cs:22,1.15844692670437)
--(axis cs:22,1.16952743226999);

\path [draw=color1, ultra thick]
(axis cs:24,1.25102443421865)
--(axis cs:24,1.26009864270443);

\path [draw=color1, ultra thick]
(axis cs:26,1.31044535046077)
--(axis cs:26,1.31606541877);

\path [draw=color1, ultra thick]
(axis cs:28,1.35068898972524)
--(axis cs:28,1.3539843436081);

\path [draw=color1, ultra thick]
(axis cs:30,1.37274783606631)
--(axis cs:30,1.37633421521574);

\path [draw=color1, ultra thick]
(axis cs:32,1.39274185248103)
--(axis cs:32,1.40013763469846);

\path [draw=color1, ultra thick]
(axis cs:34,1.43924358587663)
--(axis cs:34,1.45117487566183);

\path [draw=color1, ultra thick]
(axis cs:36,1.56294490680844)
--(axis cs:36,1.57385970857617);

\path [draw=color1, ultra thick]
(axis cs:38,1.7406257714795)
--(axis cs:38,1.74743371569999);

\path [draw=color1, ultra thick]
(axis cs:40,1.87234818657353)
--(axis cs:40,1.87649283906749);

\path [draw=color1, ultra thick]
(axis cs:42,2.00142355262513)
--(axis cs:42,2.00612721660564);

\path [draw=color1, ultra thick]
(axis cs:44,2.16749116390188)
--(axis cs:44,2.17265806686735);

\path [draw=black, ultra thick]
(axis cs:2,0.997443271534855)
--(axis cs:2,1.00116847205489);

\path [draw=black, ultra thick]
(axis cs:4,1.00009014828342)
--(axis cs:4,1.00266695428069);

\path [draw=black, ultra thick]
(axis cs:6,1.00093005912287)
--(axis cs:6,1.00357858190277);

\path [draw=black, ultra thick]
(axis cs:8,0.997379447186441)
--(axis cs:8,0.999335424608431);

\path [draw=black, ultra thick]
(axis cs:10,0.996070523648072)
--(axis cs:10,0.998325373787825);

\path [draw=black, ultra thick]
(axis cs:12,0.997446949841401)
--(axis cs:12,0.999374793748342);

\path [draw=black, ultra thick]
(axis cs:14,0.996790180714062)
--(axis cs:14,0.998291229542348);

\path [draw=black, ultra thick]
(axis cs:16,0.997910274155255)
--(axis cs:16,0.999303033537053);

\path [draw=black, ultra thick]
(axis cs:18,0.997013874003747)
--(axis cs:18,0.998432023432151);

\path [draw=black, ultra thick]
(axis cs:20,1.10687391536014)
--(axis cs:20,1.12329880258858);

\path [draw=black, ultra thick]
(axis cs:22,1.27176107340675)
--(axis cs:22,1.29801174710607);

\path [draw=black, ultra thick]
(axis cs:24,1.4082589270393)
--(axis cs:24,1.42880825244788);

\path [draw=black, ultra thick]
(axis cs:26,1.49424473218912)
--(axis cs:26,1.50634244729806);

\path [draw=black, ultra thick]
(axis cs:28,1.5533298539383)
--(axis cs:28,1.56222937683093);

\path [draw=black, ultra thick]
(axis cs:30,1.59949713481573)
--(axis cs:30,1.60537671133811);

\path [draw=black, ultra thick]
(axis cs:32,1.65650029604735)
--(axis cs:32,1.6635035501065);

\path [draw=black, ultra thick]
(axis cs:34,1.78546854542712)
--(axis cs:34,1.80035555713698);

\path [draw=black, ultra thick]
(axis cs:36,2.1975180889657)
--(axis cs:36,2.2366178084702);

\path [draw=black, ultra thick]
(axis cs:38,2.87622670044186)
--(axis cs:38,2.91705124827609);

\path [draw=black, ultra thick]
(axis cs:40,3.42997539820962)
--(axis cs:40,3.46269896076474);

\path [draw=black, ultra thick]
(axis cs:42,4.50164291563817)
--(axis cs:42,4.5513719561567);

\path [draw=black, ultra thick]
(axis cs:44,9.06974509839824)
--(axis cs:44,9.56576541442228);

\addplot [ultra thick, color0, dotted]
table {%
2 0.999958487179487
4 1.00330602564103
6 1.00460253846154
8 0.998409846153846
10 0.995197435897436
12 0.997406102564103
14 0.99736382051282
16 0.998714743589744
18 0.998128717948718
20 1.18027948717949
22 1.40578564102564
24 1.58150564102564
26 1.68733179487179
28 1.76322256410256
30 1.83033282051282
32 1.9235641025641
34 2.14061487179487
36 2.86573358974359
38 4.04924820512821
40 5.01825384615385
42 7.04923948717949
44 16.4654358974359
};
\addlegendentry{$\overline{J}_{1.25}$}
\addplot [ultra thick, color1, dashed]
table {%
2 0.998653256410257
4 0.999451076923077
6 0.999906102564103
8 0.998305025641026
10 0.999198461538461
12 0.999415641025641
14 0.99771758974359
16 0.998498564102564
18 0.997317179487179
20 1.04989323076923
22 1.16398717948718
24 1.25556153846154
26 1.31325538461538
28 1.35233666666667
30 1.37454102564103
32 1.39643974358974
34 1.44520923076923
36 1.56840230769231
38 1.74402974358974
40 1.87442051282051
42 2.00377538461538
44 2.17007461538462
};
\addlegendentry{$\overline{J}_{0.75}$}
\addplot [ultra thick, black]
table {%
2 0.999305871794872
4 1.00137855128205
6 1.00225432051282
8 0.998357435897436
10 0.997197948717949
12 0.998410871794872
14 0.997540705128205
16 0.998606653846154
18 0.997722948717949
20 1.11508635897436
22 1.28488641025641
24 1.41853358974359
26 1.50029358974359
28 1.55777961538462
30 1.60243692307692
32 1.66000192307692
34 1.79291205128205
36 2.21706794871795
38 2.89663897435897
40 3.44633717948718
42 4.52650743589744
44 9.31775525641026
};
\addlegendentry{$\overline{J}$}
\end{axis}

\end{tikzpicture}}
	\vspace{-6mm}\caption{Solid line illustrates the average quadratic cost per system, i.e., $\overline{J}$. Dashed lines $\overline{J}_{0.75}$ and $\overline{J}_{1.25}$ show the average rate of stable and unstable plants, respectively. Vertical error bars represent $95\%$ confidence intervals.}\vspace{-2.5mm}
	\label{fig:cost}
\end{figure}
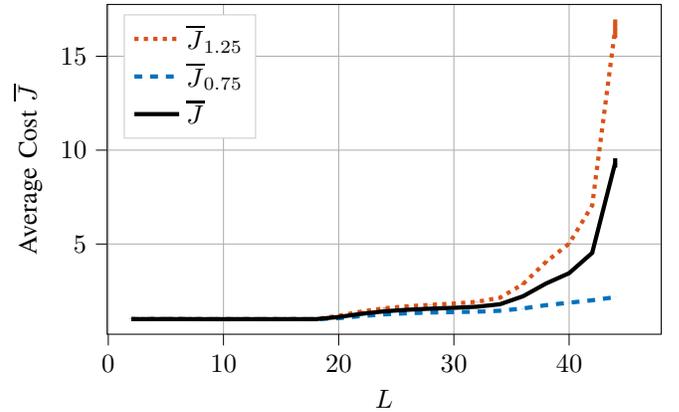

Figs. \ref{fig:rate} and \ref{fig:queueBacklog} present the interplay between the queue backlog (queued packets) at the source and the resulting average rate $\overline{r}$ admitted at the source in terms of packets per slot. In addition, the averages over all stable and unstable loops are shown. Confidence intervals for $95\%$ confidence level are included but mostly smaller than the line width in the figures. As the number of loops $L$ increases from 2 to 20, we observe that the average rate $\overline{r}$ per loop stays constant. The initial rate being equal to $1.0$ corresponds to the normalized sampling rate of control loops and implies that all of the generated packets have been successfully delivered to the respective recipient. As the number of loops exceeds 20, which is also the maximum number of transmission opportunities available in sampling period $T_k$, the CC mechanism reduces the packet admission rate. The system becomes unstable when $L\!=\!46$. From this point on, due to divergence of error in unstable loops, the CC algorithm starts to inject packets into MAC queue with full sampling rate, i.e., $\overline{r} \!=\! 1.0$, effectively overloading the network. Hence, the packet injection stops completely for the stable loops with $A^i \!=\! 0.75$, as their error norm hardly exceeds the respective threshold values. In order to avoid visual clutter, we exclude the results for $L\!=\!46$ in Figs. \ref{fig:queueBacklog}, \ref{fig:delay} and \ref{fig:cost}.

From Fig. \ref{fig:queueBacklog} we can see that together with the admission control, back-pressure solution as a MAC policy succeeds limiting queue backlog length up to 44 loops.  Since Algorithm \ref{algo:backpressure_solution} selects the user with maximum queue backlog on each link, average queue backlog stays close for both plant classes for $L \leq 44$. From the results illustrated in Fig.s \ref{fig:rate} and \ref{fig:queueBacklog}, we can conclude that the unstable loops, i.e., $A^i = 1.25$, constitute the majority of the ongoing traffic as network congestion increases. This is evident from Figure \ref{fig:delay} as well which shows the average delay in time-steps. Even though packet injection rate of the unstable loops is much higher than the stable ones, due to priorization of larger queue backlog, they are provided lower end-to-end delay. The resulting control cost is illustrated in Fig. \ref{fig:cost}. It shows that for higher number of loops, the control cost of the unstable loops is higher even though they are served higher rates and lower delays. This follows from the higher sensitivity of unstable loops to packet drops and delays \cite{zhang2001stability}. 

\glsresetall
\section{Conclusions}\label{sec:conclusions}
In this work we investigate a joint system design for NCS, from both, control and networking perspectives. We formulate the problem of minimizing the weighted sum LQG cost for stochastic LTI systems in a multi-hop network with generalized MAC layer capabilities. Optimization is performed over the set of admissible sampling, control, congestion control and scheduling strategies. We perform decomposition of the global problem into a ``primal'' control and a ``master'' networking problem, which are coupled through Lagrangian multipliers. The primal problem is solved for fixed multipliers and leads to a certainty equivalence control together with a threshold based sampling policy. A dual optimization is used for the master problem, leading to a back-pressure type scheduler with a simple pass-through congestion control. Interestingly, the Lagrange multipliers are shown to be related to queue back-logs, i.e., the local back-log could be used as ``communication price'' for the corresponding control loop. The resulting structure was applied and implemented to a two-hop cellular network, where the observations consistently verified the theoretical results.

\bibliographystyle{IEEEtran}
\bibliography{mybibliography}

\end{document}